\documentclass[12pt]{article} 
\usepackage{a41}
\usepackage{epsfig}
\usepackage{cite}
\usepackage{amsmath}
\usepackage{amssymb}
\usepackage{float}
\usepackage{verbatim}
\usepackage{amsxtra}
\usepackage{afterpage}
\usepackage[rflt]{floatflt}

\allowdisplaybreaks[4]

\newcommand\ds{\displaystyle}
\newcommand\fs{\footnotesize}
\newcommand\SH{\,\mbox{$\sqcup \! \sqcup$}\,}

\newcommand{\HA}{{\rm H}^*} 
\newcommand{\arccosh}{{\rm arccosh}}

\newcommand{\N}{\nonumber} 
\newcommand{\ep}{\varepsilon}

\newcommand{\lsim}{\raisebox{-0.07cm   }
{$\, \stackrel{<}{{\scriptstyle\sim}}\, $}}
\newcommand{\gsim}{\raisebox{-0.07cm   }
{$\, \stackrel{>}{{\scriptstyle\sim}}\, $}}

\newcommand\GeV{\,\mbox{GeV}}

\newcommand\be{\begin{eqnarray}}
\newcommand\ee{\end{eqnarray}}

\newcommand\Li{{\rm Li}}

\newcommand\Mvec{\,\mbox{\bf M}}

\allowdisplaybreaks[4]
\begin{document}
\noindent
\sloppy
\thispagestyle{empty}
\begin{flushleft}
DESY 13-063 \hfill %{\tt arXiv:1002.xxx [hep-ph]}
\\
DO-TH-14/02 \\
SFB/CPP-14-12 \\ 
LPN 14-046 \\
Higgstools 14-005\\
February 2014
\end{flushleft}
%
%\setcounter{page}{0}
% 1
%\mbox{}
\vspace*{\fill}
\hspace{-3mm}
{\begin{center}
{\Large\bfseries Calculating Massive 3-loop Graphs for Operator}

\vspace*{2mm}
{\Large\bfseries  Matrix Elements
by the Method of Hyperlogarithms}

\end{center}
}

\begin{center}
\vspace{2cm}
\large
Jakob Ablinger$^a$, Johannes Bl\"umlein$^b$, Clemens Raab$^b$, Carsten Schneider$^a$, \newline and 
Fabian 
Wi\ss{}brock$^{a,b}$\footnote{Present address: IHES, 35 Route de Chartres, 91440 Bures-sur-Yvette,
France.}

\vspace{5mm}

\vspace{5mm}
\normalsize {\itshape $^a$~Research
Institute for Symbolic Computation (RISC),\\ Johannes Kepler
University, Altenbergerstra\ss{}e 69, A-4040 Linz, Austria}
\\ 

\vspace{5mm}
\normalsize
{\itshape $^b$~Deutsches Elektronen--Synchrotron, DESY,\\
Platanenallee 6, D--15738 Zeuthen, Germany}
%\today
\end{center}

\vspace*{\fill} %
%%%%%%%%%%%%%%%%%%%%%%%%%%%%%%%%%%%%%%%%%%%%%%%%%%%%%%%%%%%%%%%%%%%%%%%%
\begin{abstract}
\noindent
We calculate convergent 3-loop Feynman diagrams containing a single massive loop 
equipped with twist $\tau =2$ local operator insertions corresponding to spin
$N$. They contribute to the massive operator matrix elements in QCD describing
the massive Wilson coefficients for deep-inelastic scattering at large virtualities.
Diagrams of this kind can be computed using an extended version to the method 
of hyperlogarithms, originally being designed for massless Feynman diagrams without
operators. The method is applied to Benz- and $V$-type graphs, belonging to the
genuine 3-loop topologies. In case of the $V$-type graphs with five massive 
propagators new types of nested sums and iterated integrals emerge. The sums
are given in terms of finite binomially and inverse binomially weighted generalized
cyclotomic sums, while the 1-dimensionally iterated integrals are based on 
a set of $\sim 30$ square-root valued letters. We also derive the asymptotic
representations of the nested sums and present the solution for $N \in \mathbb{C}$.
Integrals with a power-like divergence in $N$--space $\propto a^N, a \in \mathbb{R}, 
a > 1,$ for large values of $N$ emerge. They still possess a representation in $x$--space, 
which is given in terms of root-valued iterated integrals in the present case. The method 
of hyperlogarithms is also used to calculate higher moments for crossed box graphs with
different operator insertions.
\end{abstract}
%%%%%%%%%%%%%%%%%%%%%%%%%%%%%%%%%%%%%%%%%%%%%%%%%%%%%%%%%%%%%%%%%%%%%%%%

\vspace*{\fill} 

\newpage
%%%%%%%%%%%%%%%%%%%%%%%%%%%%%%%%%%%%%%%%%%%%%%%%%%%%%%%%%%%%%%%%%%%%%%%
%           Introduction
%%%%%%%%%%%%%%%%%%%%%%%%%%%%%%%%%%%%%%%%%%%%%%%%%%%%%%%%%%%%%%%%%%%%%%%
\section{Introduction}
\label{sec:1}
\renewcommand{\theequation}{\thesection.\arabic{equation}}
\setcounter{equation}{0} 
%%%%%%%%%%%%%%%%%%%%%%%%%%%%%%%%%%%%%%%%%%%%%%%%%%%%%%%%%%%%%%%%%%%%%%%

\vspace*{1mm}
\noindent
Massive on-shell operator matrix elements (OMEs) occur in the calculation of the
Wilson coefficients in deeply-inelastic scattering, describing these
quantities at large enough virtualities $Q^2 \gg m^2$ together with the massless
Wilson coefficients \cite{Buza:1995ie}. These OMEs are loop corrections 
to local composite operators being placed in graphs with massless external lines, which are on-shell.
Their scale is set by the mass of an internal closed fermion line. Starting at
3-loop order, graphs with more than a single mass contribute 
\cite{Ablinger:2011pb,Ablinger:2012qj}. The scale $Q^2$ from which on the asymptotic 
representation was found to apply at 2-loop order at the 1\% level for the structure 
function $F_2(x,Q^2)$ is $Q^2/m^2 \gsim 10$, with $m$ the heavy quark mass, cf. \cite{Buza:1995ie}. 
Here the asymptotic result was compared to the complete one \cite{NLO} also containing non-universal
power corrections. For $F_2(x,Q^2)$ this is a very acceptable kinematic range at HERA 
in case of $m = m_{\rm charm}$ since at lower virtualities $Q^2 \lsim 20~\GeV^2$ still
significant higher twist terms contribute \cite{HT,Alekhin:2012ig,Alekhin:2013nda}.

Beyond NLO all massive OMEs have been calculated for a series of moments $N = 10,(12,14)$ 
in the single mass case \cite{Bierenbaum:2009mv,Blumlein:2009rg} for $F_2,F_L$ and transversity, 
and the moments $N = 2,4,6$ for the contributions with two different masses 
\cite{Ablinger:2011pb,Ablinger:2012qj,BW1} 
for $F_2$ at NNLO. With these results also all contributions to the unpolarized 3-loop anomalous dimensions 
$\propto T_F$ were calculated independently for these moments and confirmed earlier results, cf. \cite{MOM}. 

In case of the massive OMEs and Wilson coefficients at general values of the Mellin variable 
$N$ all logarithmic contributions are available \cite{Bierenbaum:2010jp}, to which also the 
2-loop terms \cite{Buza:1995ie,Bierenbaum:2007qe} up to $O(\varepsilon)$ 
\cite{Bierenbaum:2008yu} contribute.\footnote{The asymptotic heavy flavor contributions to 
$F_L(x,Q^2)$ at NNLO were calculated in \cite{Blumlein:2006mh}. They, however, apply only at much 
higher virtualities than those for $F_2(x,Q^2)$.}
All $O(T_F^2 N_F)$ contributions were computed in
\cite{Ablinger:2010ty,Blumlein:2012vq}. This includes the two complete massive 3-loop OMEs
$A_{qq,Q}^{(3),\sf PS}$ and $A_{gg,Q}^{(3)}$, out of eight. Very recently also the OMEs $A_{gq}^{(3)}, 
A_{qq,Q}^{(3),\sf NS}$ and $A_{Qq}^{(3),\sf PS}$ were calculated \cite{Ablinger:2014lka}.
There are first results on the 
$T_F^2$--terms
in the equal mass case \cite{Ablinger:2011pb,Ablinger:2012ej,ABHRS13}. In the polarized case
the massive OMEs were computed to 2-loop order in Refs.~\cite{Buza:1996xr,Bierenbaum:2007pn}.
In the calculation of these diagram classes the Feynman parameter integrals are reduced to multiply 
nested finite and infinite sums \cite{Blumlein:2009ta,Blumlein:2010zv}, using representations through 
hypergeometric functions
and their generalizations \cite{HYP} and Mellin-Barnes representations \cite{MB}. The sums obtained
are then calculated using the packages {\tt Sigma} \cite{SIGMA}, {\tt EvaluateMultiSums}  and {\tt
SumProduction} \cite{EVALUATEMULTISUMS}, applying also the algebraic and structural properties of 
harmonic sums \cite{Vermaseren:1998uu,Blumlein:1998if,Blumlein:2003gb,Blumlein:2009ta,Blumlein:2009fz}, 
their associated polylogarithms \cite{Remiddi:1999ew} ,
and special constants \cite{Blumlein:2009cf}, including extensions to the cyclotomic 
\cite{Ablinger:2011te} 
and generalized harmonic sum case \cite{Moch:2001zr,Ablinger:2013cf}. These relations are encoded in 
the package {\tt HarmonicSums}, cf.~\cite{Ablinger:2013cf,Ablinger:2010kw,Ablinger:2013hcp}.\footnote{For
recent surveys on mathematical structures in zero- and single Feynman integrals in Quantum Field Theories see
\cite{SURV}.}

Beyond the above topologies at the 3-loop level also ladder and Benz-type\footnote{These graphs 
received their name being of similar form as the Mercedes-Benz symbol {\tt http://de.wikipedia.\newline
org/wiki/Mercedes-Stern}.}, $V$-type and crossed box graphs
contribute. In Ref.~\cite{Ablinger:2012qm} we calculated diagrams of the
3-loop ladder topology of up to six massive propagators, including the most demanding cases. Not all 
of these graphs could be calculated using the above technologies. 

In case the corresponding graph exhibits no poles in the dimensional parameter $\varepsilon = 
D-4$, the method of hyperlogarithms has been devised for massless 2-point topologies with an
off-shell external momentum in scalar field theory in Ref.~\cite{Brown:2008um}. This method allows
to transform the Feynman-parameter integrals into special numbers, which are for the first loop 
orders linear combinations of multiple zeta values \cite{Blumlein:2009cf} and are given in terms 
of hyperlogarithms at unity argument. The integration is organized as a consecutive mapping 
into hyperlogarithms due to the linear structure of Feynman-parameters in these integrals within 
{\it every} integration step,
which is being kept during the integration process. 
In the present paper we generalize this method
allowing for local operator insertions. Furthermore, we consider the case of massive diagrams in 
which a higher nesting of Feynman-parameters is generally expected if compared to the massless 
case. 
I.e. the formalism may lead to structures beyond linearity at an earlier stage than in the 
massless case. The local operator insertions introduce a new degree of freedom, the Mellin variable 
$N$. The corresponding Feynman diagrams are given in terms of sum-representations. In most simple cases 
harmonic sums emerge. More involved cases lead to generalized sums over rational alphabets, and also 
nested cyclotomic 
and binomial sums, as will be shown below. Interestingly, for fixed integer values of $N$ the 
corresponding graphs evaluate to rational numbers, weighted by multiple zeta values for the loop-level 
considered in this paper, similar to the case in the original approach \cite{Brown:2008um}. One may 
calculate moments up to $N = 9$ even for the most complicated 3--loop graphs 
which emerge in the present physics project. These moments can be checked 
by very different methods based on the codes {\tt MATAD} \cite{Steinhauser:2000ry}
and {\tt qexp} \cite{QEXP} at lower values of $N$. The method works, since the numerator functions are 
polynomials in the
Feynman parameters at fixed values of $N$. Partial fractioning may be performed until one obtains denominator functions 
only. However, the above number of moments is usually still far too low to try the reconstruction
of the general $N$ behaviour using the method described in \cite{Blumlein:2009tj}.

Introducing an auxiliary parameter $x$, the local operator insertions may, however, be resummed such 
that a generating function is obtained, which is expressed in terms of hyperlogarithms 
$L_{\vec{a}}(x)$. In turn the $N$th Taylor-coefficient of this function has to be obtained 
analytically. The last step can be performed in some cases using {\tt HarmonicSums} directly. In more
complex situations associated difference equations of larger order have to be established and 
solved using {\tt Sigma} \cite{SIGMA}.

The paper is organized as follows. We describe the extension of the method \cite{Brown:2008um} to 
massive operator matrix elements at 3--loop order in the presence of local operator insertions in 
Section~\ref{sec:2}. The method is applied to convergent
Benz-type and related diagrams in Section~\ref{sec:2b}, also discussing practical aspects. 
Here we also derive
the asymptotic representations of the individual graphs, which is necessary for their 
representation
for complex values of $N$ needed to perform the Mellin inversion in practical applications 
\cite{Blumlein:2009ta,ANCONT}. In Section~\ref{sec:v} we calculate graphs of the 3-loop 
$V$-topology 
with five massive propagators. They may be considered to emerge from either a ladder- or the crossed 
box-topology by removing one line. While in the former case conventional structures are obtained, in the latter case new 
nested sum-types emerge, which contain weights due to binomials of the type $\binom{2i}{i}$ both in 
the numerator and denominator. In the calculation root-valued structures in the auxiliary parameter occur in 
the last step which are responsible for these new hypergeometric terms. Aspects of the Mellin-inversion of
the contributions from binomially weighted nested sums are discussed in Section~\ref{sec:an}.
In Section~\ref{sec:c} we apply the algorithm to calculate three crossed  box-topologies for fixed integer values 
of $N$ to demonstrate the applicability of the present algorithm also for these diagrams.  Section~\ref{sec:con} 
contains the conclusions.
%%%%%%%%%%%%%%%%%%%%%%%%%%%%%%%%%%%%%%%%%%%%%%%%%%%%%%%%%%%%%%%%%%%%%%%
%           The Formalism
%%%%%%%%%%%%%%%%%%%%%%%%%%%%%%%%%%%%%%%%%%%%%%%%%%%%%%%%%%%%%%%%%%%%%%%
\section{The Formalism}
\label{sec:2}
\renewcommand{\theequation}{\thesection.\arabic{equation}}
\setcounter{equation}{0} 
%%%%%%%%%%%%%%%%%%%%%%%%%%%%%%%%%%%%%%%%%%%%%%%%%%%%%%%%%%%%%%%%%%%%%%%

\vspace*{1mm}
\noindent
We consider massive Feynman diagrams at $l=3$ loops with operator insertions in $D=4+\ep$ dimensions. 
One may represent the Feynman parameter integral $I_G$ of a graph $G$ in terms of Schwinger 
parameters 
using Symanzik \cite{SYMANZIK} or Kirchhoff polynomials \cite{KIRCHHOFF}, cf.~\cite{Bogner:2010kv}. The 
Feynman rules are given in \cite{YND,Bierenbaum:2009mv}, including those for 
the operator insertions. The integral is given by~:
%----------------------------------------------------------------------------------------------------------
\begin{eqnarray}
I_G &=& \frac{\Gamma\left(a-l D/2\right)}{\prod_j \Gamma\left(a_j\right)} \int_{0}^{\infty} 
\frac{\prod_j \alpha_j^{a_j-1} \mathit{OP}_i\left(\alpha_i,N\right)}{\Psi_G^{D/2} {M}_G^{a-l D/2}} 
\delta\left(1-\sum_{l \in v} \alpha_l~\right) d\alpha_i~.
\label{eq:IG1}
\end{eqnarray} 
%----------------------------------------------------------------------------------------------------------
Here $a_i$ denote the powers of the different propagators, $a=\sum_{i \in edges} a_i$. According to the 
Cheng-Wu theorem \cite{CHENGWU} the sum of Schwinger parameters over an arbitrary subset of edges $E$ 
in $G$ may be set equal to one, as expressed by the $\delta$-distribution in (\ref{eq:IG1}).
We associate to the graph $G$ the graph $\tilde{G}$ which is obtained by closing the external lines.
While $M_G$ 
is given by the sum of all Schwinger parameters which are attached to a massive line, the graph polynomial 
$\Psi_G$ and 
the operator insertion $\mathit{OP}_i\left(\alpha_i,N\right)$ obey the following graph theoretical 
descriptions.

For a graph with $n_v$ vertices and $n_e$ edges we define the $n_e \times n_v$ graph incidence matrix 
%----------------------------------------------------------------------------------------------------------
\begin{eqnarray}
\left(\varepsilon\right)_{e,v}&=&
\begin{cases}
1, &\text{if the edge $e$ starts at vertex $v$}\\
-1, &\text{if the edge $e$ ends at vertex $v$}\\
0,  &\text{if the edge $e$ is not connected to vertex $v$}~. 
\label{matrix1}
\end{cases}
\end{eqnarray}
%----------------------------------------------------------------------------------------------------------
We choose $\varepsilon_G$ as the matrix $n_e \times (n_v-1)$-matrix obtained from (\ref{matrix1}) 
by removing 
one arbitrary column. $\varepsilon_G$ is thus not uniquely defined and depends on the direction of the edges 
and the choice of the removed column. The graph matrix ${\sf M}_G$ reads
%----------------------------------------------------------------------------------------------------------
\begin{eqnarray}
 {\sf M}_G=
 \begin{pmatrix}
 \alpha_1 &&&&\vline&\\
 &\ddots&&&\vline& \varepsilon_G\\
 &&\ddots&&\vline&\\
 &&&\alpha_{n_e}&\vline&\\
 \hline
 &&&&\vline&\\
 &&{-~^T\varepsilon}_G&&\vline &0
 \end{pmatrix}
\end{eqnarray}
%----------------------------------------------------------------------------------------------------------

%----------------------------------------------------------------------------------------------------------
\begin{figure}
\begin{center}
\includegraphics[scale=0.9]{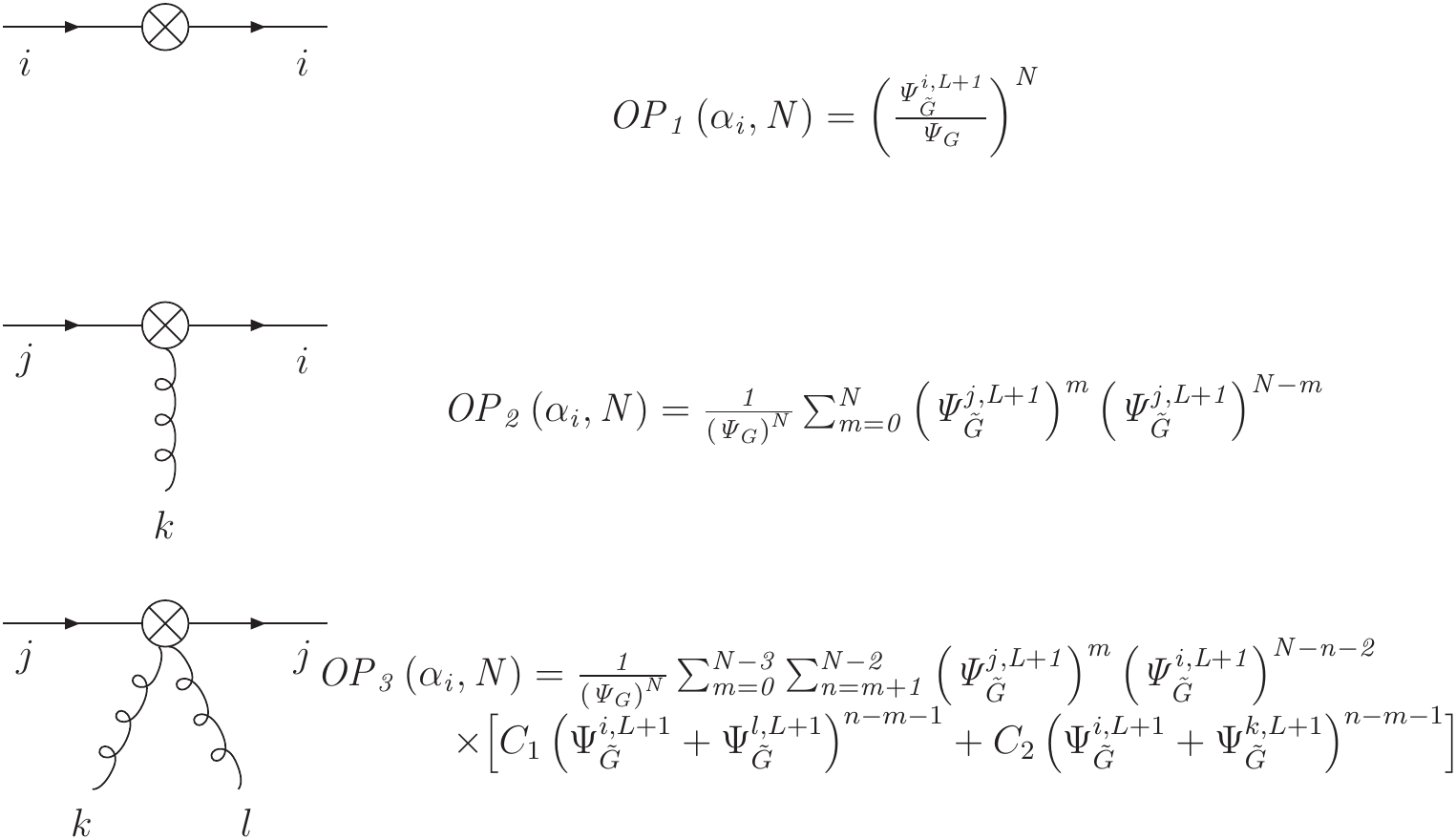}
\end{center}
\caption[]{\sf The operators expressed in terms of the graph-polynomial $\Psi_G$ and different Dodgson 
polynomials \cite{DODGSON,Brown:2009ta,Brown:2010bw} of the graph $\tilde{G}$.
\label{operators}}
\end{figure}
%----------------------------------------------------------------------------------------------------------
\noindent
The first graph polynomial $\Psi_G$ is given by $\Psi_G=-\operatorname{det}({\sf M}_G)$. Although the 
matrix 
${\sf M}_G$ is not uniquely defined $\Psi_G$, is 
independent of the possible choices for ${\sf M}_G$. If $I$,$J$,$K$ are sets of edges in the graph $G$ 
and $I$ and $J$ are of equal length, $|I|=|J|$, 
we define the Dodgson polynomials by
%----------------------------------------------------------------------------------------------------------
\begin{eqnarray}
\Psi^{I,J}_{G,K}=\pm \operatorname{det} {\sf M}_G  \left(I,J\right)\vline_{\alpha_e=0~~\forall~ 
e\in K}~,
\end{eqnarray}
%----------------------------------------------------------------------------------------------------------
with ${\sf M}_G \left(I,J\right)$ being the matrix ${\sf M}_G$ after removing all rows corresponding to 
the 
edges in $I$ and all columns corresponding to the edges in $J$. If $K$ is empty we omit it and write 
$\Psi^{I,J}_{G}$. The different operator insertions used in the present paper are expressed in terms of 
Dodgson polynomials given in Figure~\ref{operators} for the examples studied in the present paper.
The Dodgson polynomials 
$\Psi^{I,J}_{G,K}$ are only defined up to a sign, which generally depends on the orientation of the edges 
in $\varepsilon_G$ and also on the column which has been removed to define ${\sf M}_G$. 
For the present paper we were able to choose $\Psi^{I,J}_{G,K}=\operatorname{det} {\sf M}_G  
\left(I,J\right)\vline_{\alpha_e=0~~ \forall~ e\in K}$ if the directions of the edges correspond to 
the Feynman rules of Refs.~\cite{YND,Bierenbaum:2009mv}.

Under certain conditions, Feynman parameter integrals, being convergent in $D=4$ dimensions, can be cast 
into a linear combination of hyperlogaritms  $L(\vec{a},z)$ \cite{Poincare1884,LAPPO,CHEN}. In the 
following we will outline the corresponding formalism, extending the algorithm \cite{Brown:2008um}, 
given originally for massless Feynman diagrams to those with also massive lines and local operator 
insertions.
   
Let $\sigma$ be a set of 
distinct points in $\mathbb{C}$ and $\mathfrak{A}=\{a_0,a_1,...,a_N\}$ an alphabet. We form 
words described by $\vec{a}$ out of the elements of $\mathfrak{A}$, 
where each letter corresponds to an element in $\sigma$. The elements in $\sigma$ may 
be constants or rational functions of further parameters. The hyperlogarithms are defined by
%-------------------------------------------------------------------------------------------------------------
\begin{eqnarray}
L(\vec{a},z) : \mathbb{C}\setminus \sigma \rightarrow \mathbb{C}
\end{eqnarray}
%-------------------------------------------------------------------------------------------------------------
with
%-------------------------------------------------------------------------------------------------------------
\begin{eqnarray}
L(\emptyset,z) &=& 1 
\\
L(\{\underbrace{0,\cdots,0}_{\text{n times}}\},z)&=&=\frac 1 {n!} \log(z)^n
\\
L(\{a_1\},z)&=&\int_0^z dz_1 \frac 1 {z_1-a_1} \label{logs}\\
L(\{b,\vec{a}\},z) &=& \int_0^z dz_1 \frac {1} {z_1-b} L(\{\vec{a}\},z_1)~. 
\label{hlog}
\end{eqnarray}
%-------------------------------------------------------------------------------------------------------------
Here $\{...\}$ denotes an ordered set.
The weight {\sf w} of a hyperlogarithm is given by the number of letters
in $\vec{a}$. The hyperlogarithms satisfy shuffle relations, cf.~e.g.~\cite{Blumlein:2003gb},
%-------------------------------------------------------------------------------------------------------------
\begin{eqnarray}
L\left(\vec{a}_1, z\right) 
L\left(\vec{a}_2, z\right)
=
L\left(\vec{a}_1, z\right) 
\SH
L\left(\vec{a}_2, z\right)~.
\label{shuffle}
\end{eqnarray}
%-------------------------------------------------------------------------------------------------------------
In the shuffled index set one sums over all hyperlogarithms with indices such that the relative order of the 
indices in $\vec{a}_1$ and $\vec{a}_2$ is preserved. An example is given by
%-------------------------------------------------------------------------------------------------------------
\begin{eqnarray}
 L\left(\{a,b\},z\right) L\left(\{c,d\},z\right)&=& L\left(\{a,b,c,d\},z\right)+L\left(\{a,c,b,d\},z\right)
  +L\left(\{a,c,d,b\},z\right)
  \N\\&&
  +L\left(\{c,a,b,d\},z\right)+L\left(\{c,a,d,b\},z\right)+L\left(\{c,d,a,b\},z\right).
\nonumber\\
\end{eqnarray}
%-------------------------------------------------------------------------------------------------------------
The derivatives w.r.t.  the argument $z$ is  
%-------------------------------------------------------------------------------------------------------------
\begin{eqnarray}
\frac {d} {dz} L(\{b,\vec{a}\},z)&=& \frac {1} {z-b} L(\{\vec{a}\},z).
\end{eqnarray}
%-------------------------------------------------------------------------------------------------------------

The general tactic is to treat the inner most integral first and to transform the integrals
from inside to outside in terms of hyperlogarithms.  Let us start with the inner most integral
and turn to the construction of the antiderivative (primitive functions) of products of rational 
functions $R(z) = N(z)/D(z)$  and hyperlogarithms. Here we will assume that $D(z)$ factors linearly,
i.e. $D(z) = \prod_k (z - a_k)^{l_k}, l_k \in \mathbb{N}$. If there exists one integration order for 
a graph $G$ for which this property is found in each integration step such a graph is called to be 
linear reducible. The consecutive decomposition of the multiple integral into a sequence of these 
steps is called Fubini sequence. Whether or not this decomposition exists can be checked a priori 
with reduction algorithms given in Refs.~\cite{Brown:2008um,Brown:2009ta} by which also the requested order 
of integration is delivered.
Applying the shuffle relation and partial fractioning one arrives at expressions of the form
%-------------------------------------------------------------------------------------------------------------
\begin{eqnarray}
I(b,n)&=&\int dx (x+b)^n L\left(\{a_1,\vec{a}\},x\right)~.
\end{eqnarray}
%-------------------------------------------------------------------------------------------------------------
For $n=-1$ again the hyperlogarithm $L\left(\{-b,a_1,\vec{a}\},x\right)~$ is obtained.
Otherwise one applies integration by parts 
%-------------------------------------------------------------------------------------------------------------
\begin{eqnarray}
I(b,n)=\frac{(x+b)^{n+1}}{n+1} L\left(\{a_1,\vec{a}\},x \right)
-\int dx (x+b)^{n+1} \frac 1 {(n+1) (x-a_1)} L\left(\{\vec{a}\},x \right)~,
\end{eqnarray}
%-------------------------------------------------------------------------------------------------------------
where in  the last term the weight of the hyperlogarithm is reduced by one.
Applying this technique recursively, all integrals can be written in terms of hyperlogarithms that 
have to be
evaluated at its integration bounds in the $\alpha$-representation (i.e. at $0$ and $\infty$). The 
challenge is now to perform 
this evaluation, more precisely to calculate the limits. To accomplish this task, we actually 
calculate the series expansion at $0$ and at $\infty$ and express the result
again in terms of hyperlogarithms afterwards. This finally enables one to apply the presented method for the 
next integral.

Next we consider series expansions of the hyperlogarithms around $z = 0$ and 
for $z \rightarrow \infty$. 
A hyperlogarithm of weight {\sf w}  satisfies series 
representations of the form
%-------------------------------------------------------------------------------------------------------------
\begin{eqnarray}
L(\{a_1,\cdots,a_n\},z)&=&\sum_{i=0}^{\infty} \sum_{j=0}^{w} c^{(0)}_{i,j}
\log^j (z) z^i~.\\
L(\{a_1,\cdots,a_n\},z)&=&\sum_{i=0}^{\infty} \sum_{j=0}^{w} c^{(\infty)}_{i,j}
\log^j (z) z^{-i} \label{serinf}~.
\end{eqnarray}
%-------------------------------------------------------------------------------------------------------------
Following \cite{Brown:2008um} it is suitable to define the restricted regularization 
${\rm RReg}_{z\rightarrow\{0,\infty\}}$ given by the constant part of the generalized series expansion
%-------------------------------------------------------------------------------------------------------------
\begin{eqnarray}
{\rm RReg}_{z\rightarrow0} ~L(\{a_1,\cdots,a_n\},z) &=& c^{(0)}_{0,0}=0\\
{\rm RReg}_{z\rightarrow\infty}~ L(\{a_1,\cdots,a_n\},z) &=& c^{(\infty)}_{0,0} ~.
\end{eqnarray}
%-------------------------------------------------------------------------------------------------------------
One may regularize an integral by
%-------------------------------------------------------------------------------------------------------------
\begin{eqnarray}
\int_{{\rm Reg}\left(0\right)}^z f(y) dy &:=& F(z) - {\rm RReg}_{y\rightarrow0} F(y)~.
\end{eqnarray}
%-------------------------------------------------------------------------------------------------------------
The series expansions are constructed as follows. One first differentiates w.r.t. the argument of the
hyperlogarithm and then performs the series expansion of the derivative, which is of lower weight.
After this the antiderivative is calculated and the respective integration constants are fixed.
We denote the series operator by ${\rm Ser}_{y\rightarrow\infty}^{(k)}$, up to terms of
$O\left(y^{-k} \log^w(y)\right)$. One obtains
%-------------------------------------------------------------------------------------------------------------
\begin{eqnarray}
{\rm Ser}_{z\rightarrow \infty}^{(k)} L\left(\{a_1,\vec{a}\},z\right)&=&
\int_{{\rm Reg}\left(0\right)}^{z} {\rm Ser}_{z\rightarrow \infty}^{(k+1)} 
\frac{d} {dz} 
L\left(\{a_1,\vec{a}\},z\right)
+ {\rm RReg}_{z\rightarrow\infty}  L\left(\{a_1,\vec{a}\},z\right)
\nonumber\\
&=&
\int_{{\rm Reg}\left(0\right)}^{z} {\rm Ser}_{z\rightarrow \infty}^{(k+1)} 
\frac{1} {z-a_1} 
L\left(\{\vec{a}\},z\right) 
+ {\rm RReg}_{z\rightarrow\infty}  L\left(\{a_1,\vec{a}\},z\right)
~.
\end{eqnarray}
%-------------------------------------------------------------------------------------------------------------
For example, one finds
%-------------------------------------------------------------------------------------------------------------
\begin{eqnarray}
{\rm Ser}_{y\rightarrow\infty}^{(4)}\frac {d} {dz} L(\{a_1\},z) &=& \frac 1
z-\frac{a_1}{z^2}+\frac{a_1^2}{z^3}-\frac{a_1^3}{z^4}+O\left(\frac{1}{z^5}\right),
\\
{\rm Ser}_{y\rightarrow\infty}^{(3)} L(\{a_1\},z)&=&
c^{(\infty)}_{0,0}\left(\{a_1\}\right)+L(\{0\},z)-\frac {a_1} {z} -\frac {a_1^2} {2 z^2}
-\frac {a_1^3} {3 z^3} +O\left(\frac {1}{z^4}\right)~.
\end{eqnarray}
%-------------------------------------------------------------------------------------------------------------
The same method is applied to construct the series representations for hyperlogarithms 
of higher weight.\footnote{Algorithms to obtain closed forms for these expansions are known 
and have been implemented into the computer
algebra package HarmonicSums \cite{Ablinger:2013cf,Ablinger:2010kw,Ablinger:2013hcp}.} 

We now line out how the integration constants can be transformed, which is necessary in
the applications. Derivatives for a the variable $t$ of which the letters $a_i(t)$ in 
the index-set of the hyperlogarithms may depend, are computed as follows~:
%-------------------------------------------------------------------------------------------------------------
\begin{eqnarray}
\frac {\partial} {\partial t } L(\{a_{1}(t),a_{2}(t),\cdots,a_{n}(t)\},z)&=&
\int \limits_{{\rm Reg}\left(0\right)}^z dz_1 \int \limits_{{\rm Reg}(0)}^{z_{1}} dz_{2}
\cdots \int \limits_{\rm{Reg}\left(0\right)}^{z_{n-1}} dz_n 
\frac{\partial}{\partial t} \prod_{i=1}^n  \frac{1} {z_i-a_i(t)}~. 
\label{innerdiff}
\end{eqnarray}
%-------------------------------------------------------------------------------------------------------------
Note that taking the derivative with respect to the argument or an inner variable of the 
hyperlogarithm always yields expressions  which contain only hyperlogarithms of a lower 
weight. To prepare the next integration step, the constants 
%-------------------------------------------------------------------------------------------------------------
\begin{eqnarray}
c^{(\infty)}_{0,0}\left(\{a_1,\cdots,a_n\}\right)=
{\rm RReg}_{y\rightarrow\infty} L\left(\{a_1,\cdots,a_n\},y\right)
\end{eqnarray}
%-------------------------------------------------------------------------------------------------------------
have to be rewritten in terms of hyperlogarithms, such that the next integration variable 
does not appear in the respective index set. This is done by differentiating, rewriting the 
now weight-reduced expression and then forming the antiderivative again.
Let us consider the example
%-------------------------------------------------------------------------------------------------------------
\begin{eqnarray}
c^{(\infty)}_{0,0}\left(-x,-1\right) = {\rm RReg}_{y\rightarrow\infty}~L(\{-x,-1\},y)~.
\end{eqnarray}
%-------------------------------------------------------------------------------------------------------------
With
%-------------------------------------------------------------------------------------------------------------
\begin{eqnarray}
{\rm RReg}_{y\rightarrow\infty} ~ \frac {\partial}{\partial x} 
L\left(\{-x,-1\},y\right)&=& 
{\rm RReg}_{y\rightarrow\infty}~ \frac{L(\{-x\},y)}{x-1}-\frac{(y+1) L(\{-1\},y)}{(x-1) 
(x+y)}\nonumber\\
&=& -\frac{L(\{0\},x)}{x-1}
\end{eqnarray}
%-------------------------------------------------------------------------------------------------------------
one obtains
%-------------------------------------------------------------------------------------------------------------
\begin{eqnarray}
c^{(\infty)}_{0,0}\left(-x,-1\right)
&=&\int_0^x~ {\rm RReg}_{y\rightarrow\infty} ~ \frac {\partial}{\partial x'} 
L\left(\{-x',-1\},y\right) \N\\&&
+ {\rm RReg}_{x\rightarrow 0} {\rm RReg}_{y\rightarrow\infty} ~ L\left(\{-x,-1\},y\right)
\nonumber\\
&=& \int_0^x~dx~\left[ -\frac{L(\{0\},x')}{x'-1}\right] + {\rm RReg}_{y\rightarrow\infty} ~ 
L\left(\{0,-1\},y\right)
\nonumber\\
&=&-L(\{1,0\},x)+\zeta_2~,
\end{eqnarray}
%-------------------------------------------------------------------------------------------------------------
with $\zeta_k = \sum_{l=1}^\infty 1/l^k,~k \in \mathbb{N}, k \geq 2$ the Riemann 
$\zeta$-function. Special care has to be taken when evaluating constants 
like $c^{(\infty)}_{0,0}\left(a_1,\cdots,a_n\right)$ which contain letters of the form 
$x^{-i} f(x)$ with $f(x)\neq 0$ as $x\rightarrow 0$ or trailing letters of the 
form $x^i f(x)$, with  $\lim_{x \rightarrow 0} f(x)$ being finite.
In all other cases ${\rm RReg}_{x\rightarrow0} L\left({a_1,\cdots,a_n},y\right)$ is just 
obtained by taking the limit $x\rightarrow0$ under the integral. 
In the first case the limit $x\rightarrow0$ does not commute with  $y\rightarrow\infty$.  
If a hyperlogarithm does not have any trailing zero in its index set, we may substitute the 
integration variables $z_i\rightarrow a z_i$ in (\ref{hlog}) 
to obtain
%-------------------------------------------------------------------------------------------------------------
\begin{eqnarray}
 L\left(\{a_1,\cdots,a_n\},z\right)&=&L\left(\{a a_1,\cdots,a a_n\},a z\right)~. 
\label{substintvar} 
\end{eqnarray}
%-------------------------------------------------------------------------------------------------------------
In other cases trailing zeros have to be removed by means of the shuffle algebra first, 
e.g.,
%-------------------------------------------------------------------------------------------------------------
\begin{eqnarray}
L\left(\{a_1,0,0\},z\right)&=& L\left(\{a_1\},z\right) L\left(\{0,0\},z\right)
-L\left(\{0\},z\right) L\left(\{0,a_1\},z\right)
\nonumber\\ &&
+L(\{0,0,a_1\},z)
\\
&=&
L\left(\{0,0\},a\right) L\left(\{a a_1\},a z\right)-L\left(\{0\},a\right) L\left(\{a a_1,0\},a z\right)
\N\\&&
+L\left(\{a a_1,0,0\},a z\right)~, \label{substintvar0}
\end{eqnarray}
%-------------------------------------------------------------------------------------------------------------
after using the relations (\ref{substintvar}), (\ref{logs}) and (\ref{shuffle}).
Applying (\ref{substintvar}) resp. (\ref{substintvar0}) one obtains~:
%-------------------------------------------------------------------------------------------------------------
\begin{eqnarray}
c^{(\infty)}_{0,0}\left(\{x^{-i} f_1(x),\cdots,f_n(x)\}\right)&=& 
{\rm RReg}_{y\rightarrow\infty} L\left(\{x^{-i} f_1(x),\cdots,f_n(x)\},y\right)
\N\\
&=& {\rm RReg}_{y\rightarrow\infty} L\left(\{f_1(x),\cdots, x^{i} f_n(x)\},y x^{i}\right)
\N\\
&=& {\rm RReg}_{y\rightarrow\infty} \Biggl[{\rm Ser}_{z\rightarrow\infty}^{(0)} 
L\left(\{f_1(x),\cdots, x^{i} f_n(x)\},z\right)\Biggr] \vert_{z=y x^{i}}
\N\\
&=& \Biggl[{\rm Ser}_{z\rightarrow\infty}^{(0)} L\left(\{f_1(x),\cdots, x^{i} 
f_n(x)\},z\right)\Biggr] \vert_{z=x^{i}}~.
\end{eqnarray}
%-------------------------------------------------------------------------------------------------------------
By definition ${\rm Ser}_{z\rightarrow\infty}^{(0)} L\left(\{f_1(x),\cdots, x^{i} 
f_n(x)\},z\right)$ does depend on the variable $z=y x^{i}$ only logarithmically and the
operation ${\rm RReg}_{y\rightarrow\infty}$ in the second last step is easily performed.
In the case of trailing letters of the type $x^i f(x)$ with $f(x)$ finite as $x\rightarrow0$, 
the limit $x\rightarrow0$ does not commute with the implicit limits contained in the definition 
of the hyperlogarithm. Here we apply the identity
%-------------------------------------------------------------------------------------------------------------
{\small
\begin{eqnarray}
{\rm RReg}_{x\rightarrow0} L\left(\{x^{i_1} f_1(x),\cdots,x^{i_n} f_n(x)\},y\right)&=&
{\rm RReg}_{x\rightarrow0} L\left(\{x^{i_1-1} f_1(x),\cdots,x^{i_n-1} 
f_n(x)\},\frac{y}{x}\right)
\nonumber\\
&=& {\rm Ser}_{y\rightarrow\infty}^{(0)} {\rm RReg}_{x\rightarrow0} L\left(\{x^{i_1-1} 
f_1(x),\cdots,x^{i_n-1} f_n(x)\},y\right)
\label{RRegA}
\nonumber\\ 
\end{eqnarray}
}
%-------------------------------------------------------------------------------------------------------------

\noindent
on the parts containing the respective letters as many times as needed. The identity 
(\ref{RRegA}) is derived by considering the change of integrations variables $z_i\rightarrow 
\frac{z_i'}{x}$ in (\ref{hlog}).
We illustrate this in the following example~:
%-------------------------------------------------------------------------------------------------------------
{\small
\begin{eqnarray}
{\rm RReg}_{x\rightarrow 0} L\left(\{-2,-\frac{x}{2}\},y\right)
&=&
{\rm RReg}_{x\rightarrow 0} \int_0^{y} \frac{dz_1}{z_1+2} {\rm RReg}_{x\rightarrow 0} 
L\left(\{-\frac{1}{2}\},\frac{z_1}{x}\right)
\N\\&=&
\int_0^{y} \frac{dz_1}{z_1+2} \Biggl[{\rm Ser}_{z_1\rightarrow \infty}^{(0)} 
L\left(\{-\frac{1}{2}\},z_1\right)\Biggr]
\N\\&=&
\int_0^{y} \frac{dz_1}{z_1+2} \Biggl[\ln 2 + L\left(\{0\},z_1\right)\Biggr]
\N\\&=&
L\left(\{-2\},y\right) \ln 2 + L\left(\{-2,0\},y\right)~.
\end{eqnarray}
}
%-------------------------------------------------------------------------------------------------------------

\noindent
The previous steps are repeated for all further integration variables until we have rewritten 
all constants in a way suitable for the following parametric integrations.

That far we have described the algorithm for a finite loop diagram built of propagators and 
vertices for a renormalizable quantum field theory. The present application is more general
as also local operator insertions shall be dealt with. A consistent set of Feynman rules
in case of Quantum Chromodynamics has been presented in Ref.~\cite{Bierenbaum:2009mv}. As a 
consequence of the light-cone expansion \cite{LCE} the local operator insertions emerge as 
polynomials of degree $N$, $N \in \mathbb{N}$, as has been outlined above. For any integer 
value the present formalism can be applied through which the moments of the corresponding OME
are obtained. With growing values of $N$ both the requested CPU time and memory to perform this
computation will grow significantly, usually with a nearly constant factor by going from $N
\rightarrow N+2$. All finite 3--loop topologies can be dealt with this method up to a certain 
moment, i.e. the present method is equivalent for finite diagrams to {\tt MATAD} 
\cite{Steinhauser:2000ry}, which, however, can handle divergent graphs as well. In 
Section~\ref{sec:c} we will illustrate this for the most complicated graphs in the present project.

To use the present method also in case of general values of the Mellin variable $N$, the 
following resummation into a generating function in the parameter $t$
of the operator-polynomials is applied, cf.~\cite{Ablinger:2012qm}~:
%-------------------------------------------------------------------------------------------------------------
\begin{eqnarray}
 \mathit{OP}_i \left(\alpha_i,t\right)=\sum_{N=0}^{\infty} t^N  
\mathit{OP}_i \left(\alpha_i,N\right)~.
\end{eqnarray}
%-------------------------------------------------------------------------------------------------------------
Let us illustrate the derivation of the generating function for an operator insertion
on a 3--vertex. It is of the structure
%-------------------------------------------------------------------------------------------
\begin{eqnarray}
\sum_{k=0}^{N-1} A^{N-1-k} B^k = \frac{A^N - B^N}{A-B}.
\end{eqnarray}
%-------------------------------------------------------------------------------------------
The infinite resummation results into 
%-------------------------------------------------------------------------------------------
\begin{eqnarray}
A^N                   &\rightarrow& \sum_{k=0}^\infty t^k A^k = \frac{1}{1 - t A} \\
\frac{A^N - B^N}{A-B} 
&\rightarrow& \sum_{k=0}^\infty t^{k-1} \frac{A^k - B^k}{A-B} = \frac{1}{(1 - t A)(1 - 
t B)},~{\rm 
etc.}
\end{eqnarray} 
%-------------------------------------------------------------------------------------------
The generalization to the case of $l$-leg  operator-insertions is straightforward. It
leads to $(l-1)$-additional propagator terms, now containing also the variable $t$. In this way
structures are obtained which are in a form suitable for the above algorithm.
In case the auxiliary parameter $t$ does not destroy linearity in the consecutive 
integration
of Feynman parameters, finally a representation of the  generating functions by hyperlogarithms 
$L_{\vec{w}}(t)$ is obtained.

The following representations hold for the three different operators given in  
Figure~\ref{operators}~:
%-------------------------------------------------------------------------------------------------------------
\begin{eqnarray}
\mathit{OP}_1\left(\alpha_i,t\right) &=& \frac{\Psi_G}{\Psi_G-t \Psi^{i,L+1}_{\tilde{G}}}
\\
\mathit{OP}_2\left(\alpha_i,t\right)
%%\frac{1}{\Psi_G^N} \sum_{m=0}^N 
%%\left(\Psi^{i,L+1}_{\tilde{G}}\right)^{N-m} \left(\Psi^{j,L+1}_{\tilde{G}}\right)^{m}
%%&=&
%%\frac{\left(\Psi^{i,L+1}_{\tilde{G}}\right)^N-\left(\Psi^{j,L+1}_{\tilde{G}}\right)^N}{\Psi^{i,L+1}_{\tilde{G}}-\Psi^{j,L+1}_{\tilde{G}}}
&=&
\frac{\Psi_G^2}{\left({\Psi_G}-t \Psi^{i,L+1}_{\tilde{G}}\right) 
\left({\Psi_G}-t \Psi^{j,L+1}_{\tilde{G}}\right)}
\\
\mathit{OP}_3\left(\alpha_i,t\right)
%%\frac{1} {\left(\Psi_G\right)^N}  
%%\sum_{m=0}^{N-3}\sum_{n=m+1}^{N-2}
%%                      \left(\Psi_{\tilde{G}}^{j,L+1}\right)^{m} 
%%\left(\Psi_{\tilde{G}}^{i,L+1}\right)^{N-n-2}
%%                      \N\\&&
%%               \times \Bigl[ C_1  
%%\left(\Psi_{\tilde{G}}^{i,L+1}+\Psi_{\tilde{G}}^{l,L+1}\right)^{n-m-1}
%%                     +C_2 
%%\left(\Psi_{\tilde{G}}^{i,L+1}+\Psi_{\tilde{G}}^{k,L+1}\right)^{n-m-1}\Bigr]
&=&
                      \frac{\Psi_G^3}{\left({\Psi_G}-t \Psi_{\tilde{G}}^{i,L+1}\right) \left({\Psi_G}-t \Psi_{\tilde{G}}^{j,L+1}\right)}
                     \N\\&&\times
                     \Biggl[ C_1 
\frac{1}{\Psi_G-t \left(\Psi_{\tilde{G}}^{i,L+1}+\Psi_{\tilde{G}}^{l,L+1}\right)}
+C_2 
\frac{1}{\Psi_G-t \left(\Psi_{\tilde{G}}^{i,L+1}+\Psi_{\tilde{G}}^{k,L+1}\right)}\Biggr]~.
\end{eqnarray}
%-------------------------------------------------------------------------------------------------------------
The solution for the general Mellin variable $N$ can finally be obtained by calculating the $N$th 
expansion coefficient of the generating function. This usually requires to solve associated 
difference equations. Respective algorithms are encoded in the packages {\tt Sigma} 
\cite{SIGMA}, {\tt EvaluateMultiSums}, {\tt SumProduction} \cite{EVALUATEMULTISUMS}  
and {\tt HarmonicSums} \cite{Ablinger:2013cf,Ablinger:2010kw,Ablinger:2013hcp}. We finally would like to note that 
for a {\it fixed} value 
of $N$ all massive $3$-loop QCD two-point topologies turned out to be  
linear reducible in the case of a single mass scale $m$. If we introduce generating functions this changes drastically. Some diagrams 
remain linear reducible, others can be transformed into linear reducible diagrams via a 
variable transformation. There are, however, also cases for which no sequence could be found to restore linear 
reducibility.

%-------------------------------------------------------------------------------------------------------------
\begin{figure}
\begin{center}
 \includegraphics{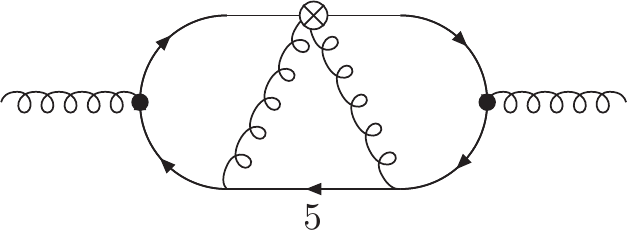}
 \end{center}
 \caption{\sf 
A $V$-topology diagram. 
\label{Diag5}}
\end{figure}
%-------------------------------------------------------------------------------------------------------------

One of the finite diagrams we would like to calculate is the scalar graph shown in 
Figure~\ref{Diag5} in Section~\ref{sec:v}. For this diagram  no completely linear reducible 
integration order exists a priori.\footnote{A corresponding remark in 
Ref.~\cite{Panzer:2014gra} is incorrect.} 
The linearization of some quadratic forms occurring can be performed introducing 
complex letters. A final quadratic form appears in the last step only and can be 
dealt with remapping the tracing variable to gain linear reducibility by
%-------------------------------------------------------------------------------------------------------------
\begin{eqnarray}
\int_0^{\infty} dy \frac{L\left(\{\cdots\},y\right)} {y^2+y (2+t)+1} 
&=& \int_0^{\infty} dy \frac{L\left(\{\cdots\},y\right)}
{\left(y+1+t/2+\sqrt{t^2+4 t}/2\right) \left(y+1+t/2-\sqrt{t^2+4 t}/2 \right)}
\nonumber\\ 
\end{eqnarray}
%-------------------------------------------------------------------------------------------------------------
Applying the transformation $t={4 x^2}/(1-x^2)$ yields
%-------------------------------------------------------------------------------------------------------------
\begin{eqnarray}
&& \int_0^{\infty} dy \left(x^2-1\right)^2 \frac{L\left(\{\cdots\},y\right)}
  {\left(y (x^2-1)-1-3 x^2 +2 x\right) \left(y (x^2-1)-1-3 x^2 -2 x\right)}.
\end{eqnarray} 
%------------------------------------------------------------------------------------------------------------- As 
The final expression will consist of hyperlogarithms in the new variable $x=\sqrt{{t}/{t+4}}$. 
More 
evolved techniques have to be applied to obtain the $N$-space representation, see Section~\ref{sec:an}. 

We now 
turn to the calculation of specific finite 3-loop topologies applying the above methods. 
%%%%%%%%%%%%%%%%%%%%%%%%%%%%%%%%%%%%%%%%%%%%%%%%%%%%%%%%%%%%%%%%%%%%%%% 
% Benz-Graphs 
%%%%%%%%%%%%%%%%%%%%%%%%%%%%%%%%%%%%%%%%%%%%%%%%%%%%%%%%%%%%%%%%%%%%%%% 
\section{Benz-Graphs} \label{sec:2b} 
\renewcommand{\theequation}{\thesection.\arabic{equation}} \setcounter{equation}{0} 
%%%%%%%%%%%%%%%%%%%%%%%%%%%%%%%%%%%%%%%%%%%%%%%%%%%%%%%%%%%%%%%%%%%%%%% 

\begin{figure}[H] 
\centering \includegraphics[scale=1.2]{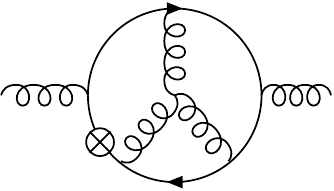} 
\caption{\sf The 3-loop Benz diagram for $I_1(N)$, 
Eq.~(\ref{EX1a}).} \label{benzA} 
\end{figure}

\vspace*{1mm}
\noindent
Let us first consider so-called Benz 
topologies. A first example is given in Figure~\ref{benzA}. 
%------------------------------------------------------------------------------------------- 
%------------------------------------------------------------------------------------------- 
\noindent 
Here all 
powers of the propagators are chosen as $\nu_i = 1$. Using the method described in Section~2
one obtains the following expression~: 
%------------------------------------------------------------------------------------------- 
\begin{eqnarray} 
\hat{I}_1(x)&=& \frac{1} {(1+N) (2+N) x} \Biggl\{ \Bigl[2 L_{-1}(x) -2 (-1+2 x) L_1(x) - 4 L_{1,1}(x) \Bigr] 
\zeta_3 
\nonumber \\ && - 3 L_{-1,0,0,1}(x) + 2 L_{-1,0,1,1}(x) - 2 x L_{0,0,1,1}(x) + 3 x L_{0,1,0,1}(x) \N\\ && -x 
L_{0,1,1,1}(x) + (-3 + 2 x) L_{1,0,0,1}(x) + 2 x L_{1,0,1,1}(x) - L_{1,0,1,1,1}(x) \N\\ && - (5 x-1) 
L_{1,1,0,1}(x) + x L_{1,1,1,1}(x), - 2 L_{1,0,0,1,1}(x) + 3 L_{1,0,1,0,1}(x) \nonumber \\ && + 2 L_{1,1,0,0,1}(x) 
+ 2 L_{1,1,0,1,1}(x) - 5 L_{1,1,1,0,1}(x) + L_{1,1,1,1,1}(x) \Biggr\}~. \label{EX1} 
\end{eqnarray} 
%------------------------------------------------------------------------------------------- 
Here the global $N$-dependent factors stem from pre-manufacturing.
The 
hyperlogarithms in (\ref{EX1}) are even harmonic polylogarithms (HPLs) over the alphabet $\{0,1,-1\}$ 
\cite{Remiddi:1999ew}. 
Considering (\ref{EX1}) as a power series in $x$, the $N$th coefficient
of this expression in $x$ has to be extracted analytically in order 
to recover the original integral. This can be achieved using the {\tt {GetMoment}} function of the package {\tt 
HarmonicSums}, cf. \cite{Ablinger:2013cf}. One may also use guessing-methods to obtain the corresponding 
difference equation based on a huge number of moments, cf.~\cite{Blumlein:2009tj}, and obtain the $N$th 
coefficient by solving this equation using {\tt Sigma} \cite{SIGMA}.

The $N$th Taylor coefficient of (\ref{EX1}) is given as the following
representation in harmonic sums~:
%-------------------------------------------------------------------------------------------
\begin{eqnarray}
\label{EX1a}
I_1(N)&=&
\frac{1}{(N+1) (N+2) (N+3)}
\Biggl\{
\frac{P_1}{(1+N)^3 (2+N)^3 (3+N)^3}
\nonumber \\ &&
-\frac{2 \left(-1+(-1)^N+N+(-1)^N N\right)}{(1+N)} \zeta_3
-(-1)^N S_{-3}
-\frac{N}{6 (1+N)} S_1^3
+\frac{1}{24} S_1^4
-\frac{1}{4} S_4
\nonumber \\ &&
-\frac{\left(7+22 N+10 N^2\right)}{2 (1+N)^2 (2+N)} S_2
-\frac{19}{8} S_2^2
-\frac{1+4 N+2 N^2}{2 (1+N)^2(2+N)} S_1^2
+\frac{9}{4} S_2 S_1^2
-\frac{(-9+4 N)}{3 (1+N)} S_3
\nonumber \\ &&
-{2 (-1)^N } S_{-2,1}
+\frac{(-1+6 N)}{(1+N)} S_{2,1}
+\frac{P_2}{(1+N)^3
    (2+N)^2 (3+N)^2} S_1
\nonumber \\ &&
+{4} \zeta_3 S_1
-\frac{(-2+7 N)}{2 (1+N)} S_2 S_1
+\frac{13}{3} S_3 S_1
-{7} S_{2,1} S_1
-{7} S_{3,1}
+{10} S_{2,1,1}
\Biggr\},
\end{eqnarray}
%-------------------------------------------------------------------------------------------
with
%-------------------------------------------------------------------------------------------
\begin{eqnarray}
P_1(N) &=& 648+1512 N+1458 N^2+744 N^3+212 N^4+32 N^5+2 N^6
\\
P_2(N) &=& 54+207 N+246 N^2+130 N^3+32 N^4+3 N^5~.
\end{eqnarray}
%-------------------------------------------------------------------------------------------
The harmonic sums are denoted by \cite{Vermaseren:1998uu,Blumlein:1998if}
%-------------------------------------------------------------------------------------------
\begin{eqnarray}
S_{b,\vec{a}}(N) = \sum_{k=1}^N \frac{({\rm sign}(b))^k}{k^{|b|}} S_{\vec{a}}(k),~~~~S_\emptyset = 1~, b, a_i \in \mathbb{Z} \backslash 
\{0\}
\end{eqnarray}
%-------------------------------------------------------------------------------------------
and we use the short-hand notation $S_{\vec{a}}(N) \equiv S_{\vec{a}}$.
For all finite sum structures one easily derives the recursive shift relation
%-------------------------------------------------------------------------------------------
\begin{eqnarray}
I_1(N+1) = I_1(N) + F_1(N)~.
\label{eqSHIFT}
\end{eqnarray}
%-------------------------------------------------------------------------------------------
All harmonic sums can be written in terms of polynomial factors in $S_1(N)$ and those 
\cite{Blumlein:2009ta}, which have representations by factorial series \cite{FACT}. 
The singularities of these sums are located at the non-positive integers, implying that these 
are meromorphic functions. Furthermore the physical expressions may exhibit singularities due 
to rational factors. The rightmost singularity is determined by the spin of the particles involved. 
In case of massless spin--1 (1/2, 0) particles singularities up to $N = 1~(0,-1)$ can occur. The 
asymptotic representation of both types of sums can be uniquely determined and is automated by the 
code {\tt HarmonicSums}. The asymptotic representation and the shift-relation (\ref{eqSHIFT})
allow the analytic continuation of integrals like $I_1(N)$ into the complex plane.
The uniqueness of the analytic continuation can be proven by an extension of Carlson's theorem
\cite{Ablinger:2013cf}. It is carried out either from the even {\it or} odd integers $N$ in
the sum expression, depending on the crossing relations of the process described, cf.~\cite{LCE}. Therefore
alternating sums and factors $(-1)^N$ have a definite meaning prior to the analytic continuation
$N \in \mathbb{C}$. 

For Eq.~(\ref{EX1a}) the asymptotic expansion is given by 
%-------------------------------------------------------------------------------------------
\begin{eqnarray}
I_1^{\rm asy}(N) &\simeq& 
%--
       \left(
              \frac{1}{24 N^3}
              -\frac{1}{4 N^4}
              +\frac{25}{24 N^5}
              -\frac{15}{4 N^6}
              +\frac{301}{24 N^7}
              -\frac{161}{4 N^8}
              +\frac{3025}{24 N^9}
-\frac{1555}{4 N^{10}}
\right) \ln^4(\bar{N}) 
%--
\nonumber\\ &&
       +\Bigl(
             -\frac{1}{6 N^3}
             +\frac{5}{4 N^4}
             -\frac{421}{72 N^5}
             +\frac{45}{2 N^6}
             -\frac{18803}{240 N^7}
             +\frac{10313}{40 N^8}
             -\frac{2480627}{3024 N^9}
\nonumber\\ &&
+\frac{1288247}{504 N^{10}}
\Bigr) \ln^3(\bar{N}) 
%--
+\Bigl(
      -\frac{3}{2 N^4}
      +\frac{551}{48 N^5}
      -\frac{2699}{48 N^6}
      +\frac{652013}{2880 N^7}
      -\frac{98339}{120 N^8}
      +\frac{2805553}{1008 N^9}
\nonumber\\ &&
      -\frac{290543}{32 N^{10}}
\Bigr) 
      \ln^2(\bar{N}) 
%--
      +\Bigl(
             -\frac{11}{2 N^4}
             +\frac{947}{24 N^5}
             -\frac{8887}{48 N^6}
             +\frac{103891}{144 N^7}
             -\frac{36580757}{14400 N^8}
\nonumber\\ &&
             +\frac{2181959741}{259200 N^9}
-\frac{11373443593}{423360 N^{10}}
\Bigr) 
             \ln(\bar{N}) 
%--
       -\frac{16}{N^4}
       +\frac{2713}{24 N^5}
       -\frac{14114}{27 N^6}
       +\frac{773389}{384 N^7}
\nonumber\\ && 
      -\frac{152225303}{21600 N^8}
       +\frac{12096164219}{518400 N^9}
       -\frac{4428508717429}{59270400 N^{10}}
\nonumber\\ &&
%--
      +\zeta_2
      \Biggl[\left(
                  \frac{9}{4 N^3}
                  -\frac{27}{2 N^4}
                  +\frac{225}{4 N^5} 
                  -\frac{405}{2 N^6}
                  +\frac{2709}{4 N^7}
                  -\frac{4347}{2 N^8}
                  +\frac{27225}{4 N^9}
-\frac{41985}{2 N^{10}}
\right) \ln^2(\bar{N}) \nonumber\\ &&
%--
      +\left(
      -\frac{7}{2 N^3}
      +\frac{111}{4 N^4}
      -\frac{1063}{8 N^5}
      +\frac{1035}{2 N^6}
      -\frac{145147}{80 N^7}
      +\frac{239811}{40 N^8}
      -\frac{2141827}{112 N^9}
+ \frac{3342261}{56 N^{10}}
\right) \ln(\bar{N}) \nonumber\\ &&
%--   
      -\frac{7}{N^4}
      +\frac{2603}{48 N^5}
      -\frac{12755}{48 N^6}
      +\frac{340949}{320 N^7}
      -\frac{92045}{24 N^8}
      +\frac{9325513}{720 N^9}
      -\frac{28247675}{672 N^{10}}
\Biggr] \nonumber\\ &&
%--
+\zeta_3
      \Biggl[\left(
                  -\frac{17}{3 N^3}
                 +\frac{34}{N^4}
                 -\frac{425}{3 N^5}
                 +\frac{510}{N^6}
                 -\frac{5117}{3 N^7}
                 +\frac{5474}{N^8}
                -\frac{51425}{3 N^9}
+\frac{52870}{N^{10}}
\right) \ln(\bar{N}) \nonumber\\ &&
%--
       +\frac{26}{3 N^3}
       -\frac{121}{2 N^4}
       +\frac{9857}{36 N^5}
       -\frac{1035}{N^6}
       +\frac{428011}{120 N^7}
       -\frac{233281}{20 N^8}
       +\frac{55892059}{1512 N^9}
       -\frac{28953679}{252 N^{10}}
\Biggr] \nonumber\\ &&
\nonumber\\ &&
       +\zeta_2^2
\Bigr(
              \frac{241}{40 N^3}
              -\frac{723}{20 N^4}
              +\frac{1205}{8 N^5}
              -\frac{2169}{4 N^6}
              +\frac{72541}{40 N^7}
              -\frac{116403}{20 N^8}
              +\frac{145805}{8 N^9}
\nonumber\\ &&
-\frac{224853}{4 N^{10}}
\Bigl) + O\left(\frac{\ln^4(\bar{N})}{N^{11}}\right)~, 
\label{eqasy1}
\end{eqnarray}
%-------------------------------------------------------------------------------------------
with $\bar{N} = N \exp({\gamma_E})$ and $\gamma_E$ the Euler-Mascheroni constant. 

Let us now consider further topologies, exhibiting 
different levels of complexity, characterized by the type of the contributing nested 
sums.
%-------------------------------------------------------------------------------------------
\begin{figure}[H]
\centering
\includegraphics[scale=1.2]{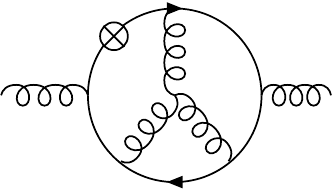}   
\caption{\sf The 3-loop Benz diagram for $I_2(N)$, Eq.~(\ref{EX1b}).}
\label{benzB}
\end{figure}
%-------------------------------------------------------------------------------------------
%%%%%%%%%%%%%%%%%%%%%%%%%%%%%%%%%%%%%%%%%%%%%%%%%%
%
%
%     QQQQQQQ
%    Q g   g Q
% ggQ   g g   Qgg
%    X   g   Q
%     Q  g  Q
%      QQQQQ
%
%
%%%%%%%%%%%%%%%%%%%%%%%%%%%%%%%%%%%%%%%%%%%%%%%%%%
Following the above algorithm, integral $I_2(N)$ defined by the graph in Figure~\ref{benzB}, yields~:
%----------------------------------------------------------------------------------------------
\begin{eqnarray}
\label{EX1b}
I_2(N)&=&
\frac {1} {(N+1) (N+2) (N+3)}
\Biggl\{
\frac{2 \left(N+3\right)}{(N+1)^3 (N+2)}
-\frac{4 \left(-4-3 N+2^{2+N}(N+1) \right)}{N+1} \zeta_3
\nonumber\\&&
+\frac{1}{2 (N+1) (N+2)} S_1^2
-\frac{1}{2} S_1^3
+\frac{\left(-1+9 N+4 N^2\right)}{2 (N+1)^2(N+2)} S_2
-\frac{5 \left(N+2 \right)}{2} S_2^2
\nonumber\\&&
- 3 S_3
-\frac{3 \left(N+2\right)}{2} S_4
-\frac{(5+3 N)}{N+1} S_{2,1}
-\frac{N^2-3}{(N+1)^3(N+2)} S_1 
+4 \left(N+2\right) S_1 \zeta_3
\nonumber\\&&
-\frac{7 }{2} S_1 S_2
- 2 \left(N+2\right) S_1 S_{2,1}
+2 \left(N+2\right)S_{3,1}
+2^{4+N} S_{1,2}\left(\frac{1}{2},1\right)
\nonumber\\&&
+ 4 \left(N+2\right) S_{2,1,1}
+2^{3+N} S_{1,1,1}\left(\frac{1}{2},1,1\right)
\Biggr\}~.
\end{eqnarray}
%-----------------------------------------------------------------------------
This integral contains generalized harmonic sums and also
terms of the $O(2^N)$, which cancel in the asymptotic expansion.
%-----------------------------------------------------------------------------
\begin{eqnarray}
I_2^{\rm asy}(N) &=&
\left(
-\frac{1}{2 N^3}
+\frac{3}{N^4}
-\frac{25}{2 N^5}
+\frac{45}{N^6}
-\frac{301}{2 N^7}
+\frac{483}{N^8}
-\frac{3025}{2 N^9}
+\frac{4665}{N^{10}}
\right)
   \ln^3(\bar{N})
\N\\ &&
%---
+\left(
-\frac{19}{4 N^4}
+\frac{297}{8 N^5}
-\frac{196}{N^6}
+\frac{72289}{80 N^7}
-\frac{163837}{40 N^8}
+\frac{6772187}{336 N^9}
-\frac{6652459}{56 N^{10}}
\right) 
\ln^2(\bar{N})
%---
\N\\ &&
+\Biggl(
-\frac{2}{N^3}
+\frac{14}{N^4}
-\frac{6089}{72 N^5}
+\frac{33071}{72 N^6}
-\frac{17131999}{7200 N^7}
+\frac{22857919}{1800 N^8}
-\frac{1113784177}{14700 N^9}
%-------------
\N\\ &&
+\frac{19063098643}{35280 N^{10}}
\Biggr) 
\ln(\bar{N})
-\frac{4}{N^3}
+\frac{35}{2 N^4}
-\frac{4181}{108 N^5}
-\frac{24331}{432 N^6}
+\frac{16232209}{12000 N^7}
\N\\ &&
-\frac{863086111}{72000 N^8}
+\frac{1575813188009}{16464000 N^9}
-\frac{483184825009}{592704 N^{10}}
%----
\N\\ &&
+   \Biggl[\left(
-\frac{7}{2 N^3}
+\frac{21}{N^4}
-\frac{175}{2 N^5}
+\frac{315}{N^6}
-\frac{2107}{2 N^7}
+\frac{3381}{N^8}
-\frac{21175}{2 N^9}
+\frac{32655}{N^{10}}
\right)
   \ln(\bar{N})
%---
\N\\ &&
+\frac{3}{N^3}
-\frac{133}{4 N^4}
+\frac{4819}{24 N^5}
-\frac{1945}{2 N^6}
+\frac{347613}{80 N^7}
-\frac{783477}{40 N^8}
+\frac{490035913}{5040 N^9}
-\frac{97672721}{168 N^{10}}
\Biggr] \zeta_2
%---
\N\\ &&
+\left(
\frac{3}{N^3}
-\frac{18}{N^4}
+\frac{75}{N^5}
-\frac{270}{N^6}
+\frac{903}{N^7}
-\frac{2898}{N^8}
+\frac{9075}{N^9}
-\frac{27990}{N^{10}}
\right) 
\zeta_3
+\Biggl(
\frac{27}{10 N^2}
\N\\
%---
&&
-\frac{54}{5 N^3}
+\frac{351}{10 N^4}
-\frac{108}{N^5}
+\frac{3267}{10 N^6}
-\frac{4914}{5 N^7}
+\frac{29511}{10 N^8}
-\frac{8856}{N^9}
+\frac{265707}{10 N^{10}}
\Biggr) \zeta_2^2 
\N\\ &&
+ O\left(\frac{\ln^3(\bar{N})}{N^{11}}\right)~.
\end{eqnarray}
%-------------------------------------------------------------------------------------------
\begin{figure}[H]
\centering
\includegraphics[scale=1.2]{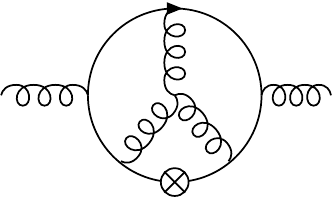}   
\caption{\sf The 3-loop Benz diagram for $I_3(N)$, Eq.~(\ref{EX1c}).}
\label{benzC}
\end{figure}
%-------------------------------------------------------------------------------------------
Diagram $I_3(N)$ differs from diagram $I_1(N)$ by moving the operator insertion to one 
propagator to the right. The result obtained is much more simple than for $I_1(N)$, cf.~(\ref{EX1a}),
and is given in terms of a few harmonic sums only,
%-----------------------------------------------------------------------------------------------
%%%%%%%%%%%%%%%%%%%%%%%%%%%%%%%%%%%%%%%%%%%%%%%%%%
%
%
%     QQQXQQQ
%    Q g   g Q
% ggQ   g g   Qgg
%    Q   g   Q
%     Q  g  Q
%      QQQQQ
%
%
%%%%%%%%%%%%%%%%%%%%%%%%%%%%%%%%%%%%%%%%%%%%%%%%%%
%-----------------------------------------------------------------------------------------------
\begin{eqnarray}
\label{EX1c}
%&& 
I_3(N) = \frac {1} {(N+1) (N+2)^2}
\Biggl\{\frac{4}{(N+1)^2 (N+2)}-\frac{4 S_1}{(N+2)}+4 S_2\Biggr\},
\end{eqnarray}
{with the asymptotic representation}
\begin{eqnarray}
I_3^{\rm asy}(N) &=&
\left(
-\frac{4}{N^4}
+\frac{28}{N^5}
-\frac{124}{N^6}
+\frac{444}{N^7}
-\frac{1404}{N^8}
+\frac{4092}{N^9}
-\frac{11260}{N^{10}}
\right)
\ln(\bar{N})
%----
\N\\ &&
-\frac{4}{N^4}
+\frac{20}{N^5}
-\frac{181}{3 N^6}
+\frac{133}{N^7}
-\frac{2009}{10 N^8}
+\frac{1297}{30 N^9}
+\frac{728377}{630 N^{10}}
%---
\N\\ &&
+\left(
\frac{4}{N^3}
-\frac{20}{N^4}
+\frac{68}{N^5}
-\frac{196}{N^6}
+\frac{516}{N^7}
-\frac{1284}{N^8}
+\frac{3076}{N^9}
-\frac{7172}{N^{10}}
\right)
\zeta_2 
+ O\left(\frac{\ln(\bar{N})}{N^{11}}\right)~.
\N\\
\end{eqnarray}
Further Benz-diagrams are shown in Figures~\ref{benzD}, \ref{benzE}.
%-----------------------------------------------------------------------------------------------
\begin{figure}[H]
\centering
\includegraphics[scale=1.2]{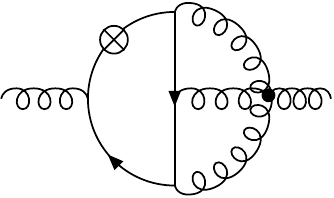}   
\caption{\sf The 3-loop Benz diagram for $I_4(N)$, Eq.~(\ref{EX1d}).}
\label{benzD}
\end{figure}
%-----------------------------------------------------------------------------------------------
%%%%%%%%%%%%%%%%%%%%%%%%%%%%%%%%%%%%%%%%%%%%%%%%%%
%
%
%      QQQgg
%     Q  Q  g
%    X   Q   g
% ggQ    Qgggggggg
%    Q   Q   g
%     Q  Q  g
%      QQQgg
%
%
%%%%%%%%%%%%%%%%%%%%%%%%%%%%%%%%%%%%%%%%%%%%%%%%%%
Integral $I_4(N)$ is given by
%-----------------------------------------------------------------------------------------------
\begin{eqnarray}
\label{EX1d}
I_4(N)&=&
\frac {1} {(N+1) (N+2)}
\Biggl\{
\frac{P_3}{(N+1) (N+2)} \zeta_3
\N\\&&
+ \frac{1}{N+2} S_{-3}
+\frac{(-1)^N}{2 (N+2)} S_1^3
-\frac{(-1)^N (3+2 N)}{2 (N+1)^2 (N+2)} S_2
+\frac{5 (-1)^N}{2} S_2^2
\N\\&&
+\frac{(-1)^N (3+2 N)}{2 (N+1)^2(N+2)} S_1^2
-\frac{(-1)^N}{2} S_2 S_1^2
+\frac{3 (-1)^N (4+3 N)}{(N+1) (N+2)} S_3
+3 (-1)^N S_4
\N\\&&
+\frac{2}{(N+2)} S_{-2,1}
+{2 (-1)^N } \zeta_3 S_1\left(2\right)
+\frac{2 (-1)^N (3+N) }{(N+1) (N+2)} S_{2,1}
-{12 (-1)^N } S_1 \zeta_3
\N\\&&
+\frac{(-1)^N (5+7 N) }{2 (N+1) (N+2)} S_1 S_2
+{3 (-1)^N } S_1 S_3
+{4 (-1)^N } S_{2,1} S_1
-{4 (-1)^N} S_{3,1}
\N\\
%\end{eqnarray}
%\begin{eqnarray}
&&
-\frac{4 \left((-1)^N 2^{2+N}-3 (-2)^N N+3 (-1)^N 2^{1+N} N\right)}{(N+1)
  (N+2)} S_{1,2}\left(\frac{1}{2},1\right)
-{5 (-1)^N } S_{2,1,1}
\N\\&&
+\frac{2 \left(-(-1)^N 2^{2+N}-13 (-2)^N N+5 (-1)^N 2^{1+N}
    N\right)}{(N+1) (N+2)} S_{1,1,1}\left(\frac{1}{2},1,1\right)
\N\\&&
-{2 (-1)^N } S_{1,1,2}\left(2,\frac{1}{2},1\right)
-{(-1)^N}  S_{1,1,1,1}\left(2,\frac{1}{2},1,1\right)
\Biggr\},\\
P_3(N) &=& 2 \left(1-13 (-1)^N+(-1)^N 2^{3+N}+N-7 (-1)^N N+3 (-1)^N
    2^{1+N}N\right), 
\end{eqnarray}
%-----------------------------------------------------------------------------------------------
containing generalized harmonic sums.

The asymptotic representation of diagram $I_4$ reads
%-----------------------------------------------------------------------------------------------
\begin{eqnarray}
I_4^{\rm asy}(N) &=& 
(-1)^N 
\Biggl\{
\Biggl[
-\frac{1793}{2 N^{10}}
+\frac{769}{2 N^9}
-\frac{321}{2 N^8}
+\frac{129}{2 N^7}
-\frac{49}{2 N^6}
+\frac{17}{2 N^5}
-\frac{5}{2 N^4}
+\frac{1}{2 N^3}
\Biggr]
\ln^3(\bar{N})
\nonumber\\ &&
+\Biggl[
-\frac{3}{2 N^3}
+\frac{21}{2 N^4}
-\frac{363}{8 N^5}
+\frac{1323}{8 N^6}
-\frac{9389}{16 N^7}
+\frac{183573}{80 N^8}
-\frac{538097}{48 N^9}
+\frac{123450851}{1680 N^{10}}
\Biggr] 
\nonumber\\ &&
\times \ln^2(\bar{N})
+\Biggl[
-\frac{7}{N^4}
+\frac{429}{8 N^5}
-\frac{6763}{24 N^6}
+\frac{662993}{480 N^7}
-\frac{3542309}{480 N^8}
+\frac{79274089}{1680 N^9}
\nonumber\\ &&
-\frac{89308307}{240 N^{10}}
\Biggr] 
\ln(\bar{N})
+
\Biggl[
\Biggl[
-\frac{1}{2 N^2}
+\frac{3}{2 N^3}
-\frac{7}{2 N^4}
+\frac{15}{2 N^5}
-\frac{31}{2 N^6}
+\frac{63}{2 N^7}
-\frac{127}{2 N^8}
+\frac{255}{2 N^9}
\nonumber\\ &&
-\frac{511}{2 N^{10}}
\Biggr]
\ln^2(\bar{N})
+\Biggl[
\frac{3}{N^3}
-\frac{203}{12 N^4}
+\frac{247}{4 N^5}
-\frac{7457}{40 N^6}
+\frac{20271}{40 N^7}
-\frac{3251987}{2520 N^8}
+\frac{528337}{168 N^9}
\nonumber\\ &&
-\frac{5348629}{720 N^{10}}
\Biggr]
\ln(\bar{N})
+\Biggl[
-\frac{5}{N^3}
+\frac{285}{8 N^4}
-\frac{3887}{24 N^5}
+\frac{181091}{288 N^6}
-\frac{1151603}{480 N^7}
+\frac{7293811}{720 N^8}
\nonumber\\ &&
-\frac{14793223}{280 N^9}
+\frac{217689527539}{604800 N^{10}}
\Biggr]
\Biggr]
\zeta_2 
+
\Biggl[
\Biggl[
-\frac{1}{N^2}
+\frac{3}{N^3}
-\frac{7}{N^4}
+\frac{15}{N^5}
-\frac{31}{N^6}
+\frac{63}{N^7}
\nonumber\\ && 
-\frac{127}{N^8}
+\frac{255}{N^9}
-\frac{511}{N^{10}}
\Biggr]
\ln(\bar{N})
+
\Biggl[
-\frac{3}{2 N^3}
+\frac{67}{12 N^4}
-\frac{59}{4 N^5}
+\frac{1363}{40 N^6}
-\frac{2949}{40 N^7}
+\frac{388153}{2520 N^8}
\nonumber\\ && 
-\frac{53027}{168 N^9}
+\frac{460691}{720 N^{10}}
\Biggr]
\Biggr]
\zeta_3 
+
\Biggl[
-\frac{12}{5 N^2}
+\frac{36}{5 N^3}
-\frac{84}{5 N^4}
+\frac{36}{N^5}
-\frac{372}{5 N^6}
+\frac{756}{5 N^7}
-\frac{1524}{5 N^8}
\nonumber\\ &&
+\frac{612}{N^9}
-\frac{6132}{5 N^{10}}
\Biggr] 
\zeta_2^2
+
\Biggl[
\frac{4}{N^3}
-\frac{49}{4 N^4}
+\frac{181}{216 N^5}
+\frac{27119}{144 N^6}
-\frac{40222139}{27000 N^7}
+\frac{1251907}{125 N^8}
\nonumber\\ &&
-\frac{10792338497459}{148176000 N^9}
+\frac{18342053050631}{29635200 N^{10}}
\Biggr]
\Biggr\}
+ O\left(\frac{\ln^3(\bar{N})}{N^{11}}\right)~.
\end{eqnarray}
%-----------------------------------------------------------------------------------------------
%-----------------------------------------------------------------------------------------------
\begin{figure}[H]
\centering
\includegraphics[scale=1.2]{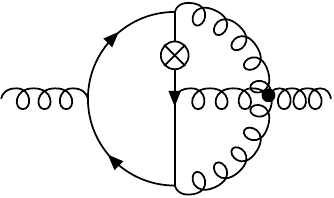}   
\caption{\sf The 3-loop Benz diagram for $I_5(N)$, Eq.~(\ref{EX1e}).}
\label{benzE}
\end{figure}
%-----------------------------------------------------------------------------------------------
Despite diagrams $I_4$ and $I_5$ are topologically quite similar, their result turns out to be structurally 
different. Integral $I_5(N)$ is given by
%-----------------------------------------------------------------------------------------------
%%%%%%%%%%%%%%%%%%%%%%%%%%%%%%%%%%%%%%%%%%%%%%%%%%
%
%
%      QQQgg
%     Q  X  g
%    Q   Q   g
% ggQ    Qgggggggg
%    Q   Q   g
%     Q  Q  g
%      QQQgg
%
%
%%%%%%%%%%%%%%%%%%%%%%%%%%%%%%%%%%%%%%%%%%%%%%%%%%
%-----------------------------------------------------------------------------------------------
\begin{eqnarray}
\label{EX1e}
I_5(N)&=&
\frac {(-1)^N} {(N+1) (N+2)}
\Biggl\{
-\frac{2 \left(2+ (-1)^N(2+N)\right)}{(1+N)} \zeta_3
+\frac{3  }{(N+1)^2} S_2
+\frac{5 }{2} S_2^2
+\frac{3 }{2} S_4
\N\\&&
+\frac{2 }{(N+1)} S_{2,1}
-\frac{2 }{(N+1)^3} S_1
-{4 } \zeta_3 S_1
+{2 } S_{2,1} S_1
-{2 } S_{3,1}
-{4 } S_{2,1,1}
\Biggr\}~,
\end{eqnarray}
%-----------------------------------------------------------------------------------------------
with the asymptotic representation
%-----------------------------------------------------------------------------------------------
\begin{eqnarray}
I_5^{\rm asy}(N) &=& (-1)^N \Biggl\{
\Biggl(
\frac{2}{N^3}
-\frac{15}{2 N^4}
+\frac{166}{9 N^5}
-\frac{445}{12 N^6}
+\frac{59153}{900 N^7}
-\frac{7987}{75 N^8}
+\frac{1185269}{7350 N^9}
\N\\ &&
-\frac{227247}{980 N^{10}}
\Biggr) \ln(\bar{N}) 
%---
+\Biggl(
-\frac{3}{N^3}
+\frac{27}{2 N^4}
-\frac{41}{N^5}
+\frac{105}{N^6}
-\frac{2449}{10 N^7}
+\frac{5397}{10 N^8}
-\frac{40158}{35 N^9}
\N\\ &&
+\frac{16686}{7 N^{10}}
\Biggr) \zeta_2
%--
+ \Biggl(
\frac{4}{N^3}
-\frac{25}{2 N^4}
+\frac{2885}{108 N^5}
-\frac{883}{18 N^6}
+\frac{381781}{4500 N^7}
-\frac{1312181}{9000 N^8}
\N\\ &&
+\frac{4756944037}{18522000 N^9}
-\frac{386004953}{823200 N^{10}}
\Biggr)
%---
+\Biggl(
-\frac{27}{10 N^2}
+\frac{81}{10 N^3}
-\frac{189}{10 N^4}
+\frac{81}{2 N^5}
-\frac{837}{10 N^6}
\N\\ &&
+\frac{1701}{10 N^7}
-\frac{3429}{10 N^8}
+\frac{1377}{2 N^9}
-\frac{13797}{10 N^{10}}
\Biggr) \zeta_2^2 \Biggr\}
%---
+\Biggl(
-\frac{2}{N^2}
+\frac{4}{N^3}
-\frac{6}{N^4}
+\frac{8}{N^5}
-\frac{10}{N^6}
\N\\ &&
+\frac{12}{N^7}
-\frac{14}{N^8}
+\frac{16}{N^9}
-\frac{18}{N^{10}}
\Biggr) \zeta_3% 
+% 
O\Biggl(\frac{\ln(\bar{N})}{N^{11}}\Biggr)~.%
\end{eqnarray}
%-----------------------------------------------------------------------------------------------
\begin{figure}[H]
\centering
\includegraphics[scale=1.2]{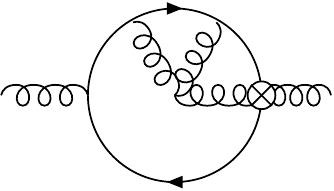}   
\caption{\sf The 3-loop Benz diagram for $I_6(N)$, Eq.~(\ref{EX1f}).}
\label{benzF}
\end{figure}
%-----------------------------------------------------------------------------------------------
Finally we consider diagram 6 as an example for convergent Benz-graphs. Applying the above algorithm
one obtains~:
%-----------------------------------------------------------------------------------------------
\begin{eqnarray}
\label{EX1f}
I_6(N)&=&
C_1 \Biggl\{
\frac{P_4}{(N+1)^5 (N+2)^5 (N+3)}
-(-1)^N \frac{P_5}{(N+1)^5 (N+2)^5 (N+3)}
+10 S_{-5}
\N\\&&
+\frac{P_6}
      {2 (N+1)^3 (N+2)^3 (N+3)^2} S_1^2
+\frac{P_7}
      {2 (N+1)^2 (N+2)^2 (N+3)^2} S_1^3
+\frac{4}{N+3} S_1 S_{-3}
\N\\&&
-\frac{P_8}
      {(N+1)^2 (N+2)^2 (N+3)^2} S_{-3}
+3 S_2 S_{-3}
+\frac{5}{N+3} S_4
-S_5
-2 S_{-4,1}
\N\\ &&
+\Biggl[\frac{3 (-1)^N P_9}
            {(N+1)^2 (N+2)^2 (N+3)^2}
+\frac{P_{10}}
      {(N+1)^2 (N+2)^2 (N+3)^2}
\Biggr] S_3
-\frac{2}{(N+3)^2} S_{-2,1}
\N\\&&
-8 S_{-2,3}
+  \Bigg[ 
(-1)^N \frac{2 P_9}
            {(N+1)^2 (N+2)^2 (N+3)^2} 
-4 S_{-2}
-4 S_2
-\frac{2 (N+2)} {N+3} S_1 \left(2\right)
\N\\&&
+\frac{2 P_{11}}
  {(N+1)^2 (N+2)^2 (N+3)^2}
+ 2^{N+2} \frac{P_{12}}
  {(N+1)^2 (N+2)^2 (N+3)^2}
\Biggr] \zeta_3
-5 S_{2,-3}
\N\\&&
+\left[-\frac{17+23 N+9 N^2+N^3}
             {(N+1) (N+2) (N+3)^2}
-(-1)^N \frac{58+84 N+43 N^2+10 N^3+N^4}
             {(N+1)^2 (N+2)^2 (N+3)^2}\right] S_{2,1}
\N\\&&
+\frac{2 \left(17+27 N+15 N^2+3 N^3\right)}
      {(N+1)^3 (N+2)^3 (N+3)} S_{-2}
-\frac{2}{N+3} S_{-2} S_2
+2 S_3 S_{-2}
+2 S_{2,1} S_{-2}
\N\\&&
- (-1)^N \frac{P_{13}}
              {(N+1)^3 (N+2)^3 (N+3)^2} S_2
+\frac{P_{14}}
      {2 (N+1)^3 (N+2)^3 (N+3)^2} S_2
\N\\&&
-S_3 S_2 -2 S_{-2,1} S_2 +2 S_{2,1} S_2
+(-1)^N \frac{2 \left(7+6 N+N^2\right) \left(9+10 N+3 N^2\right)}
             {(N+1)^4 (N+2)^4 (N+3)^2} S_1
\N\\ && + \Biggl[
\frac{P_{15}}
      {(N+1)^4 (N+2)^4 (N+3)^2}
+(-1)^N \frac{P_9}
             {(N+1)^2 (N+2)^2 (N+3)^2} S_2 \Biggr] S_1 
-2 S_{2,3}
\N\\&&
+\frac{7 \left(61+136 N+123 N^2+55 N^3+12 N^4+N^5\right)}
      {2 (N+1)^2 (N+2)^2 (N+3)^2} S_2 S_1
+\frac{5 }{N+3} S_1 S_3
\N\\ &&
-\frac{1}{N+3} S_1 S_{2,1}
-\frac{(9+N)}{N+3} S_{3,1}
+S_{4,1}
-\frac{2^{2+N} \left(4+7 N+N^2\right) }
      {(N+1) (N+3)^2} S_{1,2}\left(\frac{1}{2},1\right)
-2 S_{2,1,-2}
\N\\
&&
-2 S_{2,2,1} - \frac{2(2+N)}{3+N} S_{2,1,1}
-2 S_{3,1,1}
-\frac{2^{1+N} \left(4+7 N+N^2\right)}
      {(N+1) (N+3)^2} S_{1,1,1} \left(\frac{1}{2},1,1\right)
\N\\
&&
+\frac{2 (N+2)}{N+3} S_{1,1,2} \left(2,\frac{1}{2},1\right)
+3 S_{2,1,1,1}
+\frac{(N+2)}{N+3} S_{1,1,1,1} \left(2,\frac{1}{2},1,1\right)
\Biggr\}
\N\\
&&
+C_2 \Biggl\{-(-1)^N \frac{P_{16}}{(1+N)^5 (2+N)^5 (3+N)}
+\frac{P_{17}}{(1+N)^5 (2+N)^5 (3+N)}
-10 S_{-5}
\N\\ &&
+\frac{38+45 N+16 N^2+N^3}{(1+N)^2 (2+N)^2 (3+N)} S_{-3}
-\frac{4 S_1}{3+N} S_{-3}
-3 S_2 S_{-3} +S_5 +2 S_{-4,1} +8 S_{-2,3}
\N\\ &&
+\left[ -\frac{1}{2 (3+N)} + \frac{(-1)^N}{(2+N) (3+N)} \right] S_2^2
+S_3 S_2  +2 S_{-2,1} S_2 
+5 S_{2,-3}
\N\\&&
+  2 \Biggl[
\frac{11+15 N+7 N^2+N^3}{(1+N)^2 (2+N)^2 (3+N)}
-(-1)^N \frac{23+28 N+10 N^2+N^3}{(1+N)^2 (2+N)^2
  (3+N)}
+2 S_{-2}
\N\\ &&
+\left(-\frac{1}{3+N}-(-1)^N \frac{1}{(2+N) (3+N)}\right) S_1
+ S_2\Biggr] \zeta_3
-\frac{(-1)^N}{2 (2+N) (3+N)} S_1^2 S_2
\N\\
&&
-\frac{2 (-1)^N \left(5+6 N+2 N^2\right)}{(1+N)^2 (2+N)^3 (3+N)} S_1^2
+\frac{2}{(2+N) (3+N)} S_{-2,1}
\N\\&&
+\left(
\frac{9+10 N+3 N^2}{(1+N)^2 (2+N)^2 (3+N)} 
-\frac{3 (-1)^N \left(23+28 N+10 N^2+N^3\right)}{(1+N)^2 (2+N)^2
  (3+N)}
\right) S_3
\N\\ &&
+\left(-\frac{3}{2 (3+N)}+\frac{3 (-1)^N}{2 (2+N) (3+N)}\right) S_4
\N\\&&
-\frac{(-1)^N \left(-8-7 N+N^3\right)}{(1+N)^3 (2+N)^3 (3+N)} S_2 
+\frac{17+27 N+15 N^2+3 N^3}{(1+N)^3 (2+N)^3 (3+N)} S_2 
\N\\ &&
\N\\&&
-\frac{2 \left(17+27 N+15 N^2+3 N^3\right)}{(1+N)^3 (2+N)^3 (3+N)} S_{-2}
+\frac{2 S_2}{3+N} S_{-2}
-2 S_3 S_{-2}
-2 S_{2,1} S_{-2}
\N\\&&
+\frac{(-1)^N \left(23+28 N+10 N^2+N^3\right)}{(1+N)^2 (2+N)^2 (3+N)} S_{2,1}
\N\\ &&
-\frac{4 (-1)^N (3+2 N) \left(3+3 N+N^2\right)}{(1+N)^4 (2+N)^3 (3+N)}
S_1 
-\frac{(-1)^N \left(23+28 N+10 N^2+N^3\right)}{(1+N)^2 (2+N)^2
  (3+N)} S_1 S_2
\N\\&&
+\left(-\frac{1}{3+N}
-\frac{3 (-1)^N}{(2+N) (3+N)}\right) S_3 S_1
+\frac{(-1)^N}{(2+N) (3+N)} S_1 S_{2,1}
\N\\ &&
+\left(\frac{1}{3+N}+\frac{5 (-1)^N}{(2+N) (3+N)}\right) S_{3,1}
+2 S_{2,1,-2}
-\frac{5 (-1)^N}{(2+N) (3+N)} S_{2,1,1}
\Biggr\}~.
\label{eqI6N}
\end{eqnarray}
%-----------------------------------------------------------------------------
Here $C_1$ and $C_2$ are group-theoretic factors accounting e.g. for color degrees. We leave them unspecified 
since only scalar graphs are calculated, see Figure~1. The polynomials in (\ref{eqI6N}) read
%-----------------------------------------------------------------------------
\begin{eqnarray}
P_4(N) &=& -70-108 N-18 N^2+49 N^3+30 N^4+5 N^5
\\
P_5(N) &=& -70-104 N- 3 N^2+70 N^3+43 N^4+8 N^5
\\
P_6(N) &=& 47+98 N+81 N^2+30 N^3+4 N^4
\\
P_7(N) &=& 61+136 N+123 N^2+55 N^3+12 N^4+N^5
\\
P_8(N) &=& 112+168 N+89 N^2+18 N^3+N^4
\\
P_9(N) &=& 58+84 N+43 N^2+10 N^3+N^4
\\
P_{10}(N) &=& 48+213 N+274 N^2+150 N^3+36 N^4+3 N^5
\\
P_{11}(N) &=& -126-284 N-259 N^2-116 N^3-25 N^4-2 N^5
\\
P_{12}(N) &=& 16+60 N+80 N^2+47 N^3+12 N^4+N^5
\\
P_{13}(N) &=& 51+103 N+81 N^2+29 N^3+4 N^4
\\
P_{14}(N) &=& 325+758 N+669 N^2+262 N^3+38 N^4
\\
P_{15}(N) &=& 160+391 N+396 N^2+204 N^3+52 N^4+5 N^5
\\
P_{16}(N) &=& 142+370 N+388 N^2+203 N^3+52 N^4+5 N^5
\\
P_{17}(N) &=& 142+374 N+403 N^2+224 N^3+65 N^4+8 N^5.
\end{eqnarray}
%-----------------------------------------------------------------------------
The asymptotic expansion of $I_6$ is given by
%-----------------------------------------------------------------------------
\begin{eqnarray}
I_6^{\rm asy}(N) &=& 
C_1 
\Biggl\{
\Biggr[
\frac{1}{4 N^2}
-\frac{19}{12 N^3}
+\frac{15}{2 N^4}
-\frac{1889}{60 N^5}
+\frac{247}{2 N^6}
-\frac{38935}{84 N^7}
+\frac{3371}{2 N^8}
-\frac{359009}{60 N^9}
+\frac{41679}{2 N^{10}}
\Biggr]
\nonumber\\ &&
\times
\ln^3(\bar{N})
+\Biggl[
\frac{1}{8 N^2}
+\frac{23}{12 N^3}
-\frac{223}{12 N^4}
+\frac{45229}{400 N^5}
-\frac{280379}{480 N^6}
+\frac{66622583}{23520 N^7}
-\frac{23133233}{1680 N^8}
\nonumber\\ &&
+\frac{724473271}{10080 N^9}
-\frac{2931192779}{6720 N^{10}}
%\Biggl] 
\Biggl] 
\ln^2(\bar{N})
+\Biggl[
(-1)^N
\Biggl[
\frac{1}{N^7}
-\frac{21}{N^8}
+\frac{242}{N^9}
-\frac{1998}{N^{10}}
\Biggr]
\nonumber\\ &&
-\frac{7}{8 N^2}
+\frac{95}{18 N^3}
-\frac{3371}{288 N^4}
-\frac{69017}{2000 N^5}
+\frac{8462677}{14400 N^6}
-\frac{7789424551}{1646400 N^7}
+\frac{323933401}{9800 N^8}
\nonumber\\ &&
-\frac{247879811629}{1058400 N^9}
+\frac{3111216830509}{1693440 N^{10}}
\Biggr] 
\ln(\bar{N})
+\Biggl[
-\frac{24}{5}
+\frac{43}{10 N}
-\frac{129}{10 N^2}
+\frac{387}{10 N^3}
-\frac{1161}{10 N^4}
\nonumber\\ &&
+\frac{3483}{10 N^5}
-\frac{10449}{10 N^6}
+\frac{31347}{10 N^7}
-\frac{94041}{10 N^8}
+\frac{282123}{10 N^9}
-\frac{846369}{10 N^{10}}
\Biggr] 
\zeta_2^2
+(-1)^N 
\Biggl[
-\frac{1}{N^4}
+\frac{51}{4 N^5}
\nonumber\\ &&
-\frac{884}{9 N^6}
+\frac{14041}{24 N^7}
-\frac{1768501}{600 N^8}
+\frac{657507}{50 N^9}
-\frac{262301037}{4900 N^{10}}
\Biggr]
+\Biggl[
(-1)^N 
\Biggl[
-\frac{3}{N^3}
+\frac{27}{N^4}
-\frac{159}{N^5}
\nonumber\\ &&
+\frac{765}{N^6}
-\frac{3249}{N^7}
+\frac{12663}{N^8}
-\frac{46443}{N^9}
+\frac{163377}{N^{10}}
\Biggr]
-\frac{3}{2 N^2}
+\frac{19}{2 N^3}
-\frac{45}{N^4}
+\frac{1889}{10 N^5}
-\frac{741}{N^6}
\nonumber\\ &&
+\frac{38935}{14 N^7}
-\frac{10113}{N^8}
+\frac{359009}{10 N^9}
-\frac{125037}{N^{10}}
\Biggl] 
\zeta_3
+\zeta_2
\Biggl[
(-1)^N 
\Biggl[
-\frac{5}{2 N^4}
+\frac{295}{12 N^5}
-\frac{605}{4 N^6}
\nonumber\\ &&
+\frac{89029}{120 N^7}
-\frac{127147}{40 N^8}
+\frac{31520947}{2520 N^9}
-\frac{2616665}{56 N^{10}}
\Biggr]
+\Biggl[
(-1)^N 
\Biggl[
-\frac{1}{N^3}
+\frac{9}{N^4}
-\frac{53}{N^5}
\nonumber\\ &&
+\frac{255}{N^6}
-\frac{1083}{N^7}
+\frac{4221}{N^8}
-\frac{15481}{N^9}
+\frac{54459}{N^{10}}
\Biggr]
+\frac{7}{4 N^2}
-\frac{133}{12 N^3}
+\frac{105}{2 N^4}
-\frac{13223}{60 N^5}
+\frac{1729}{2 N^6}
\nonumber\\ &&
-\frac{38935}{12 N^7}
+\frac{23597}{2 N^8}
-\frac{2513063}{60 N^9}
+\frac{291753}{2 N^{10}}
\Biggr] 
\ln(\bar{N})
+3 \zeta_3
-\frac{11}{8 N^2}
+\frac{569}{36 N^3}
-\frac{1225}{12 N^4}
\nonumber\\ &&
+\frac{216201}{400 N^5}
-\frac{1261231}{480 N^6}
+\frac{125654423}{10080 N^7}
-\frac{306787391}{5040 N^8}
+\frac{9847032577}{30240 N^9}
-\frac{13758651023}{6720 N^{10}}
\Biggr]
\nonumber\\ &&
+\zeta_5
-\frac{31}{16 N^2}
+\frac{2153}{216 N^3}
-\frac{5735}{128 N^4}
+\frac{40340069}{180000 N^5}
-\frac{542992637}{432000 N^6}
+\frac{659641453013}{86436000 N^7}
\nonumber\\ &&
-\frac{7397109902939}{148176000 N^8}
+\frac{962090042920501}{2667168000 N^9}
-\frac{330634683598931}{111132000 N^{10}}
%\Biggr]
\Biggr\}
\nonumber\\ &&
+C_2 
\Biggl\{
(-1)^N 
\Biggl[
\frac{2}{N^3}
-\frac{15}{N^4}
+\frac{226}{3 N^5}
-\frac{950}{3 N^6}
+\frac{18049}{15 N^7}
-\frac{12859}{3 N^8}
+\frac{511284}{35 N^9}
-\frac{337628}{7 N^{10}}
\Biggr]
\nonumber\\ &&
\times
\ln^2(\bar{N})
+(-1)^N 
\Biggl[
\frac{4}{N^3}
-\frac{20}{N^4}
+\frac{581}{9 N^5}
-\frac{2879}{18 N^6}
+\frac{132043}{450 N^7}
-\frac{39521}{180 N^8}
-\frac{1617779}{1225 N^9}
\nonumber\\ &&
+\frac{41782189}{4410 N^{10}}
\Biggr]
\ln(\bar{N})
+\Biggl[
(-1)^N 
\Biggl[
-\frac{19}{10 N^2}
+\frac{19}{2 N^3}
-\frac{361}{10 N^4}
+\frac{247}{2 N^5}
-\frac{4009}{10 N^6}
+\frac{2527}{2 N^7}
\nonumber\\ &&
-\frac{39121}{10 N^8}
+\frac{23959}{2 N^9}
-\frac{364249}{10 N^{10}}
\Biggr]
-\frac{8}{5 N}
+\frac{24}{5 N^2}
-\frac{72}{5 N^3}
+\frac{216}{5 N^4}
-\frac{648}{5 N^5}
+\frac{1944}{5 N^6}
\nonumber\\ &&
-\frac{5832}{5 N^7}
+\frac{17496}{5 N^8}
-\frac{52488}{5 N^9}
+\frac{157464}{5 N^{10}}
\Biggr] 
\zeta_2^2
+(-1)^N 
\Biggl[
\frac{5}{N^3}
-\frac{227}{8 N^4}
+\frac{3259}{27 N^5}
-\frac{395983}{864 N^6}
\nonumber\\
%%\end{eqnarray}
%%\begin{eqnarray}
&&
+\frac{1296603}{800 N^7}
-\frac{488729}{90 N^8}
+\frac{64743036461}{3704400 N^9}
-\frac{40570237223}{740880 N^{10}}
\Biggr]
+\Biggl[
(-1)^N 
\Biggl[
-\frac{1}{2 N^2}
+\frac{5}{2 N^3}
\nonumber\\ &&
-\frac{19}{2 N^4}
+\frac{65}{2 N^5}
-\frac{211}{2 N^6}
+\frac{665}{2 N^7}
-\frac{2059}{2 N^8}
+\frac{6305}{2 N^9}
-\frac{19171}{2 N^{10}}
\Biggr]
\ln^2(\bar{N})
+(-1)^N 
\Biggl[
-\frac{5}{2 N^3}
\nonumber\\ &&
+\frac{175}{12 N^4}
-\frac{695}{12 N^5}
+\frac{7763}{40 N^6}
-\frac{4759}{8 N^7}
+\frac{4409821}{2520 N^8}
-\frac{2570599}{504 N^9}
+\frac{75289517}{5040 N^{10}}
\Biggl]
\ln(\bar{N})
\nonumber\\ &&
+(-1)^N 
\Biggl[
\frac{1}{2 N^3}
-\frac{55}{8 N^4}
+\frac{122}{3 N^5}
-\frac{51913}{288 N^6}
+\frac{1006891}{1440 N^7}
-\frac{45683}{18 N^8}
+\frac{14891573}{1680 N^9}
\nonumber\\ &&
-\frac{18137689541}{604800 N^{10}}
\Biggr]
+\frac{1}{N^2}
-\frac{6}{N^3}
+\frac{161}{6 N^4}
-\frac{106}{N^5}
+\frac{3901}{10 N^6}
-\frac{6849}{5 N^7}
+\frac{976349}{210 N^8}
-\frac{538172}{35 N^9}
\nonumber\\ &&
+\frac{2092661}{42 N^{10}}
\Biggr]
\zeta_2
+\Biggl[
(-1)^N 
\Biggl[
-\frac{15}{2 N^3}
+\frac{175}{4 N^4}
-\frac{695}{4 N^5}
+\frac{23289}{40 N^6}
-\frac{14277}{8 N^7}
+\frac{4409821}{840 N^8}
\nonumber\\ &&
-\frac{2570599}{168 N^9}
+\frac{75289517}{1680 N^{10}}
\Biggr]
+(-1)^N 
\Biggl[
-\frac{3}{N^2}
+\frac{15}{N^3}
-\frac{57}{N^4}
+\frac{195}{N^5}
-\frac{633}{N^6}
+\frac{1995}{N^7}
\nonumber\\ &&
-\frac{6177}{N^8}
+\frac{18915}{N^9}
-\frac{57513}{N^{10}}
\Biggr] 
\ln(\bar{N})
\Biggr] 
\zeta_3
-\frac{1}{4 N^3}
+\frac{7}{3 N^4}
-\frac{691}{48 N^5}
+\frac{3521}{48 N^6}
-\frac{13331}{40 N^7}
\nonumber\\ &&
+\frac{55991}{40 N^8}
-\frac{6987079}{1260 N^9}
+\frac{26436619}{1260 N^{10}}
\Biggr\}
+ O\left(\frac{\ln^3(\bar{N})}{N^{11}}\right)~.
\end{eqnarray}
%-----------------------------------------------------------------------------

For all the above graphs, irrespectively of their concrete representation at integer values 
of $N$, which is of different complexity, the shift relation $N \rightarrow (N-1)$ for 
$N \in \mathbb{C}$ can be established through simpler functions correspondingly, for which 
the analytic continuation has been worked out in 
Refs.~\cite{Blumlein:2009ta,Ablinger:2011te,Ablinger:2013cf}. In case of harmonic sums and 
cyclotomic harmonic sums the singularities are located at $N \in \mathbb{Z},~N < 1$. 
The rational pre-factors may induce also singularities at $N = 1$. The generalized 
harmonic sums in $I_2, I_4$ and $I_6$ have already been studied in (3.36--3.40) 
in \cite{Ablinger:2012qm} giving the corresponding Mellin representations. They partly appear together 
with the pre-factor $2^N$. As has been seen above, the corresponding asymptotic representations
of $I_2, I_4$ and $I_6$ are well behaved. We still have to determine the positions of the 
poles of these sums in the complex plane. The following integrals have to be considered:
%-----------------------------------------------------------------------------
\begin{eqnarray}
S_1(2;N) &=& \int_0^1 dx \frac{(2x)^N-1}{x-\tfrac{1}{2}} = S_1(N) + \int_1^2 dx \frac{x^N-1}{x-1}~.
\label{eqNI1}
\end{eqnarray}
%-----------------------------------------------------------------------------
The last integral in (\ref{eqNI1}) is analytic in $\mathbb{C}$ for any finite range. Thus the 
singularities of $S_1(2;N)$ are those of $S_1(N)$; the exponential growth of the sum for 
$N \rightarrow \infty$ is canceled by other terms in the integrals $I_{2,4,6}$.
The second integral in the sum
%-----------------------------------------------------------------------------
\begin{eqnarray}
S_{1,2}\left(\frac{1}{2},1;N\right) &=& \frac{5}{8} \zeta_3 + \frac{1}{2^N} \int_0^1 dx 
x^n \frac{\Li_2(1-x)}{2-x}
\label{eqNI2}
\end{eqnarray}
%-----------------------------------------------------------------------------
has a factorial series representation \cite{FACT}. 
Here $\Li_n(x) = \sum_{k=1}^\infty (x^k/k^n),~~n \geq 0$ denotes the polylogaritm.
The singularities are thus 
located at the non-positive integers. This also applies for the sum $S_{1,1,2}\left(2, 1/2 ,1;N\right)$, 
related to the integrals 
%-----------------------------------------------------------------------------
\begin{eqnarray}
\int_0^1 dx \frac{x^N-1}{1-x} {\rm H}_{-1,0,1}(1-x)~~\text{and}~~\int_0^1 dx \frac{x^N-1}{1-x} {\rm H}_{-1}(1-x)~.
\end{eqnarray}
%-----------------------------------------------------------------------------
Here ${\rm H}_{\vec{a}}(x)$ denote the harmonic polylogarithms over the alphabet $\{0, 1, -1\}$ 
\cite{Remiddi:1999ew}.
Next we consider
%-----------------------------------------------------------------------------
\begin{eqnarray}
\frac{1}{2} \int_0^1 dx \left(\frac{x}{2}\right)^N \frac{{\rm H}_{1,1}(x)}{1-\tfrac{x}{2}} &=& \sum_{l=0}^\infty
\frac{1}{2^{N+1+l}} \int_0^1 dx x^{N+l} \ln^2(1-x) \nonumber\\
&=&  2 \sum_{l=1}^\infty \frac{1}{2^{N+l}} \frac{S_{1,1}(N+l)}{N+l}~,
\label{eqNI3}
\end{eqnarray}
%-----------------------------------------------------------------------------
with $S_{1,1}(m) = [S_1^2(m)+S_2(m)]/2$. The representations of the harmonic sums \cite{Blumlein:1998if} imply 
that (\ref{eqNI3}) converges absolutely, with poles at $-(N+l) \in \mathbb{N} \backslash \{0\}$.  
%%%%%%%%%%%%%%%%%%%%%%%%%%%%%%%%%%%%%%%%%%%%%%%%%%%%%%%%%%%%%%%%%%%%%%%
\section{V-type Diagrams with Five Massive Propagators}
\label{sec:v}
\renewcommand{\theequation}{\thesection.\arabic{equation}}
\setcounter{equation}{0} 
%%%%%%%%%%%%%%%%%%%%%%%%%%%%%%%%%%%%%%%%%%%%%%%%%%%%%%%%%%%%%%%%%%%%%%%

\begin{figure}[H]
\centering
\includegraphics[scale=0.7]{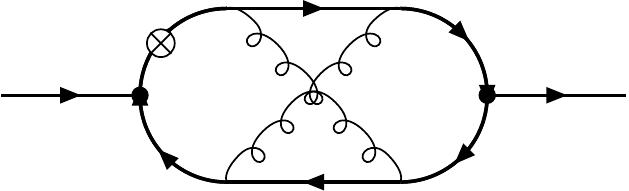}   
\includegraphics[scale=0.7]{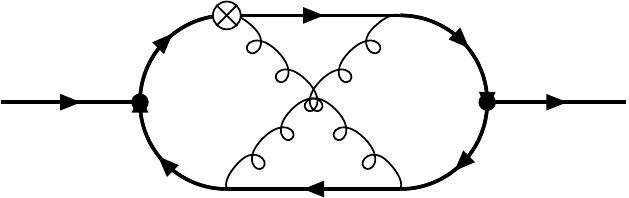}   
\includegraphics[scale=0.7]{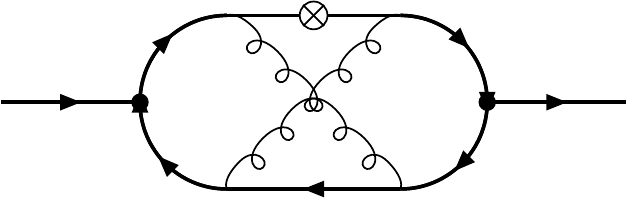}
\\
{\small
(a) \hspace*{3.9cm}
(b) \hspace*{3.9cm}
(c)} \\
\label{FIG:CB}
\caption{\sf  
Crossed-box topologies with local operator insertions.}
\end{figure}

\vspace{1mm}
\noindent
Another genuine 3--loop topology is represented by the $V$-type diagram shown in Figure~\ref{Diag5}.
According to the Feynman rules given in Figure~\ref{operators} it consists out of two contributions,
which are labeled by the constants $C_1$ and $C_2$. One may consider these terms as being obtained
by $(a)$ either expanding one line of a ladder graph or (b) the crossed box graph, cf.~Figure~\ref{FIG:CB}c, 
by applying the light-cone expansion. In the $\alpha$-representation these graphs are given by
%-----------------------------------------------------------------------------
\begin{eqnarray}
I_{7a}&=&
\int_{0}^\infty dx_1 d x_2 d \alpha_2 d\alpha_3 d\alpha_7 
\frac{\sum_{j_1=0}^N \sum_{j_2=j_1+1}^{N+1}
  \left(-T_2\right)^{j_1} \left(T_1\right)^{N+1-j_2}
  \left(T_1+T_3\right)^{j_2-j_1-1}} {U^{N+2} M} x_1 x_2 
\nonumber\\
\\
\label{eq:I5bX}
I_{7b}&=&\int_{0}^\infty dx_1 d x_2 d \alpha_2 d\alpha_3 d\alpha_7 
  \frac{\sum_{j_1=0}^N \sum_{j_2=j_1+1}^{N+1}
  \left(-T_2\right)^{j_1} \left(T_1\right)^{N+1-j_2}
  \left(T_1-T_4\right)^{j_2-j_1-1}} {U^{N+2} M} x_1 x_2,
\nonumber\\
\end{eqnarray}
%-----------------------------------------------------------------------------
where
%-----------------------------------------------------------------------------
\begin{eqnarray}
x_1&=& \alpha_1 +\alpha_6
\nonumber\\
x_2&=& \alpha_4 +\alpha_5 
\end{eqnarray}
%-----------------------------------------------------------------------------
and the different graph polynomials read
%-----------------------------------------------------------------------------
\begin{eqnarray}
M&=&x_1+x_2+\alpha_7
\nonumber\\
U &=&-\alpha_3 \alpha_2 \alpha_7-\alpha_2 \alpha_7 x_2
-\alpha_2 x_2 x_1-\alpha_3 \alpha_2 x_2-\alpha_3 \alpha_2 x_1
-\alpha_7 x_2 x_1-\alpha_3 x_2 x_1-\alpha_3 \alpha_7 x_1\nonumber\\
T_1&=&-\alpha_3 \alpha_7 \alpha_1+\alpha_3 \alpha_2 \alpha_7
-\alpha_2 \alpha_3 \alpha_1-\alpha_2 \alpha_3 \alpha_4+\alpha_2 \alpha_7 x_2
+\alpha_2 x_2 x_1-\alpha_2 x_2 \alpha_1+\alpha_3 \alpha_2 x_2
\nonumber\\ &&
+\alpha_3 \alpha_2 x_1
+\alpha_7 x_2 x_1-\alpha_7 x_2 \alpha_1+\alpha_3 x_2 x_1
-\alpha_3 x_2 \alpha_1+\alpha_3 \alpha_7 x_1
\nonumber\\
T_2&=&-(\alpha_7 \alpha_4 \alpha_2-\alpha_3 \alpha_2 \alpha_7
+\alpha_2 \alpha_3 \alpha_1+\alpha_2 \alpha_3 \alpha_4-\alpha_2 \alpha_7 x_2
-\alpha_2 x_2 x_1+\alpha_2 \alpha_4 x_1-\alpha_3 \alpha_2 x_2
\nonumber\\ &&
-\alpha_3 \alpha_2 x_1-
\alpha_7 x_2 x_1+\alpha_7 \alpha_4 x_1-\alpha_3 x_2 x_1+\alpha_3 \alpha_4 x_1
-\alpha_3 \alpha_7 x_1)
\nonumber\\
T_3&=&\alpha_7 x_2 \alpha_1+\alpha_3 x_2 \alpha_1+\alpha_3 \alpha_7 \alpha_1
-\alpha_3 \alpha_4 x_1
\nonumber\\
T_4&=&-\alpha_2 x_2 \alpha_1+\alpha_7 \alpha_4 x_1+\alpha_7 \alpha_4 \alpha_2
+\alpha_2 \alpha_4 x_1~.
\end{eqnarray}
%---------------------------------------------------------------------------
The integral $I_{7a}$, stemming from a former ladder-like topology, is expected to 
have a representation and complexity of other ladder-type diagrams considered in 
Ref.~\cite{Ablinger:2012qm} before. We first obtain the representation in terms
of hyperlogarithms~:
%---------------------------------------------------------------------------
\begin{eqnarray}
\hat{I}_{7a}(x) &=& \frac{4}{x^2 (x+1)}
\Biggl\{    - \left[L(\{0,1\},x) 
      +    L(\{0,-1\},x)\right] \zeta_3
      -  4 L(\{0,-1,-1,0,-1\},x)
\nonumber\\ &&
      -  2 L(\{0,-1,0,-1,-1\},x)
      +  2 L(\{0,-1,0, 0,-1\},x)
      +  6 L(\{0,0,-1,-1,-1\},x)
\nonumber\\ &&
      -  4 L(\{0,1,0,-1,-1\},x)
      +  2 L(\{0,1,0,0,-1\},x) \Biggl\}~.
\label{eq:5a1}
\end{eqnarray}
%---------------------------------------------------------------------------
The generating function representation is given by harmonic polylogarithms only. From (\ref{eq:5a1}) the $N$th 
Taylor coefficient 
is derived using the {\tt GetMoment} option of {\tt HarmonicSums}. $I_{7a}(N)$ is represented 
in terms of harmonic sums up weight {\sf w = 5}~:  
%---------------------------------------------------------------------------
\begin{eqnarray}
I_{7a}(N) &=&
(-1)^N \Bigg[
- \frac{12~(2 N+3)\left(N^2+3 N+3\right)}{(N+1)^3 (N+2)^3} S_1^2
+ \frac{8~\left(2 N^2+6 N+5\right)}{(N+1)^2(N+2)^2} \left[2 S_1 S_2 - S_{2,1}\right] 
\nonumber\\
&&
-\frac{8~(4 N+5)}{(N+1)^2 (N+2)^3} S_1
+8 S_3 S_2 
+ 16 S_{2,1} S_2
+ 8 S_{-2,1} S_{-2}
+8 S_5
-8S_{2,3}
+24 S_{4,1}
\nonumber\\
&&
-8 S_{-2,1,-2}
-24 S_{2,2,1}
-24 S_{3,1,1}
+\frac{4~\left(10 N^3+ 43 N^2+65 N+35\right)}{(N+1)^3 (N+2)^3} S_2
\Bigg]
\nonumber\\
&&
+\frac{8 (2 N+3) }{(N+1)^2 (N+2)^2} \left[S_{-3} - 2  S_{-2,1}\right]
\nonumber\\
&&
+ 4 \left[
(-1)^N \left(\frac{\left(2 N^2+6 N+5\right)}{(N+1)^2 (N+2)^2} 
+  S_2 + S_{-2}\right) 
-\frac{(2 N+3)}{(N+1)^2 (N+2)^2} 
\right] \zeta_3~. 
\end{eqnarray}
%---------------------------------------------------------------------------
The asymptotic representation of integral $I_{7a}$ is given by
%---------------------------------------------------------------------------
\begin{eqnarray}
\lefteqn{I_{7a}^{\rm asy}(N) \propto} \nonumber\\ &&
\Biggl[(-1)^N 
\Biggl(
-\frac{16}{N}
+\frac{40}{N^2}
-\frac{296}{3 N^3}
+\frac{240}{N^4}
-\frac{8632}{15 N^5}
+\frac{1360}{N^6}
-\frac{66536}{21 N^7}
+\frac{7280}{N^8}
-\frac{247672}{15 N^9}
+\frac{37008}{N^{10}}
\Biggr) 
\nonumber\\ &&
\times
\ln(\bar{N})
+(-1)^N 
\Biggl[
-\frac{16}{N}
-\frac{2}{N^2}
+\frac{538}{9 N^3}
-\frac{721}{3 N^4}
+\frac{18996}{25 N^5}
-\frac{6514}{3 N^6}
+\frac{12902497}{2205 N^7}
-\frac{954313}{63 N^8}
\nonumber\\ &&
+\frac{7190138}{189 N^9}
-\frac{19586179}{210 N^{10}}
+12 \zeta_3
\Biggr] \Biggr] \zeta_2 
+(-1)^N \Biggl[
        \Biggl(
 \frac{6}{N^2}
-\frac{30}{N^3}
+\frac{111}{N^4}
-\frac{360}{N^5}
+\frac{1079}{N^6}
\nonumber\\ &&
-\frac{3060}{N^7}
+\frac{8317}{N^8}
-\frac{21840}{N^9}
+\frac{278631}{5 N^{10}}
\Biggr) \ln^2(\bar{N})
+\Biggl(
 \frac{10}{N^2}
-\frac{20}{N^3}
+\frac{11}{6 N^4}
+\frac{485}{3 N^5}
-\frac{15469}{18 N^6}
\nonumber\\ &&
+\frac{19465}{6 N^7}
-\frac{13226411}{1260 N^8}
+\frac{216849}{7 N^9}
-\frac{9020336}{105 N^{10}}
\Biggr) \ln(\bar{N}) 
%---
+ \frac{1}{N^2}
+\frac{62}{3 N^3}
-\frac{7457}{72 N^4}
+\frac{31339}{90 N^5}
\nonumber\\ &&
-\frac{5369077}{5400 N^6}
+\frac{6553031}{2520 N^7}
-\frac{3416761097}{529200 N^8}
+\frac{820719223}{52920 N^9}
-\frac{192478383749}{5292000 N^{10}}
- 5 \zeta_5 \Biggr]
\nonumber\\ &&
+(-1)^N 
\Biggl(
 \frac{12}{N}
- \frac{30}{N^2}
+ \frac{74}{N^3}
- \frac{180}{N^4}
+ \frac{2158}{5 N^5}
- \frac{1020}{N^6}
+ \frac{16634}{7 N^7}
- \frac{5460}{N^8}
+ \frac{61918}{5N^9}
\nonumber\\ &&
- \frac{27756}{N^{10}}
\Biggr) \zeta_3 + O\left(\frac{\ln^2(\bar{N})}{N^{11}}\right)
\end{eqnarray}
%---------------------------------------------------------------------------
and shows a regular behaviour.

Integral $I_{7b}(N)$, related to crossed-box topologies by one additional propagator
expansion, conversely leads to new structures.

First we derive the representation of $\hat{I}_{7b}(x)$ (\ref{eq:I5bX}) in terms of iterated
integrals containing the auxiliary parameter $x$. We define
%---------------------------------------------------------------------------
\begin{eqnarray}
r = \sqrt{\frac{x}{4+x}}.
\label{eq:TRA}
\end{eqnarray}
%---------------------------------------------------------------------------
The result is given by 1405 different hyperlogarithms. The corresponding expression is too long to be given in
full form here. Instead we show a series of typical terms to illustrate different contributing 
functions~: 
%---------------------------------------------------------------------------
\begin{eqnarray}
\hat{I}_{7b}(x) &=&
-2 (3 x r+12 r-2 x) \frac{\zeta_3}{x^2 (x+1)} L(\{-1\},r)+2 (3 x r+12 r+2 x) 
\frac{\zeta_3}{x^2 (x+1)} L(\{1\},r)
\nonumber\\ &&
-\frac{23 L(\{-4,-4,-4,-4\},x)}{2 x (x+1)}
+\frac{7 L(\{-4,-4,-4,-1\},x)}{x (x+1)}+\frac{2 L(\{-4,-4,-1,-4\},x)}{x (x+1)}
\nonumber\\ &&
+\frac{9 L(\{-4,-4,-1,-1\},x)}{2 x (x+1)}+\frac{9 L(\{-4,-4,0,-4\},x)}{x (x+1)}
-\frac{8 L(\{-4,-4,0,-1\},x)}{x (x+1)}
\nonumber\\ &&
-\frac{2 L(\{-4,-4,0,1\},x)}{x (x+1)}
~~...~~
+\frac{2 (4 x r+16 r+5 x) L\big(\big\{-1,-1,0,-\frac{i}{\sqrt{3}}\big\},r\big)}{x^2 (x+1)}
\nonumber\\ &&
+\frac{2 (4 x r+16 r+5 x) L\big(\big\{-1,-1,0,\frac{i}{\sqrt{3}}\big\},r\big)}
{x^2 (x+1)}
+\frac{4 (x r+4 r-x) L\big(\big\{-1,-1,0,-\frac{1}{\sqrt{5}}\big\},r\big)}{x^2 (x+1)}
\nonumber\\ &&
+\frac{4 (x r+4 r-x) L\big(\big\{-1,-1,0,\frac{1}{\sqrt{5}}\big\},r\big)}{x^2 (x+1)}
+\frac{2 r (x+4) L(\{-1,-1,1,-1\},r)}{x^2 (x+1)}
\nonumber\\ &&
+\frac{2 r (x+4) L(\{-1,-1,1,1\},r)}
{x^2 (x+1)}
-\frac{2 r (x+4) L\big(\big\{-1,-1,1,-\frac{1}{\sqrt{5}}\big\},r\big)}{x^2 (x+1)}
\nonumber\\ &&
-\frac{2 r (x+4) L\big(\big\{-1,-1,1,\frac{1}{\sqrt{5}}\big\},r\big)}{x^2 (x+1)}
~~...
\label{eqI5b1}
\end{eqnarray}
%---------------------------------------------------------------------------
The index sets of the hyperlogarithms contain the letters
%---------------------------------------------------------------------------
\begin{eqnarray}
\left\{
1,0,-1,-4, 
\tfrac{1}{2},-\tfrac{1}{3},\tfrac{1}{3},
-\tfrac{i}{\sqrt{3}},\tfrac{i}{\sqrt{3}}
-\tfrac{1}{\sqrt{5}},\tfrac{1}{\sqrt{5}} \right\},
\label{eq:alphab}
\end{eqnarray}
%---------------------------------------------------------------------------
$x$ or $r$ as argument and reach weight {\sf w = 5}.

In the last step of integration in determining (\ref{eqI5b1}) root-valued letters appear. Both due to 
the massive case studied here and the presence of the local operator insertion in the present case no 
complete Fubini sequence is obtained in the first place. However, transformation (\ref{eq:TRA}) establishes
linear reducibility once agian and the corresponding integral can be solved.  

To derive the $N$th Taylor coefficient from (\ref{eqI5b1}) has not been straightforward.
Here we have chosen two ways. In a more simple approach we generated fixed Mellin moments
from (\ref{eqI5b1}) and used the method of guessing \cite{GUESS} to derive a corresponding 
difference equation, cf. also \cite{Blumlein:2009tj}.  We were able to generate 1500 moments. About 800 moments 
were finally needed 
to establish the difference equation. With {\tt Sigma} \cite{SIGMA} this difference equation could 
be solved in a time of 2000 seconds, through which the $N$th Taylor coefficient has been obtained. 
The method of 
guessing mostly delivers 
correct results with a failure estimated to be $\sim 10^{-60}$ \cite{GUESS}, yet it is not exact. 
Therefore we also derived from (\ref{eqI5b1}) the $N$th coefficient using {\tt Sigma} \cite{SIGMA} and {\tt 
HarmonicSums}
\cite{Ablinger:2013cf,Ablinger:2010kw,Ablinger:2013hcp}. This computation requested two days 
of computation time confirming the result obtained by the method of guessing~:
%---------------------------------------------------------------------------
\begin{eqnarray}
I_{7b} &=&
- \frac{2 (3 N+2)}{(N+1)^5 (N+2)^2}
%---
+\frac{2 \big(4 N^3+35 N^2+82 N+58\big)}{(N+1)^3 (N+2)^3} \left[S_2 + 3 S_{-2}\right]
%---
\nonumber\\ &&
-\frac{4 \big(N^3+8 N^2+23 N+20\big)}{(N+1)^2 (N+2)^2} S_3
-\frac{4 \big(N^3+8 N^2+27 N+26\big)}{(N+1)^2 (N+2)^2} S_{-3}
%---
-\frac{8 \big(N^2+6 N+7\big)}{(N+1)^2 (N+2)} S_{-2,1}
%---
\nonumber\\ &&
+2^{N+2}\frac{\big(2 N^3+12 N^2+31 N+26\big)}{(N+1)^2 (N+2)^2} 
\left[S_{1,2}\left(\frac{1}{2},1\right)
+ 3 S_{1,2}\left(\frac{1}{2},-1\right)\right]
\nonumber\\ &&
%----
+\frac{(-1)^N}{\ds \binom{2 N}{N}} 
\Biggl\{
%---
-\frac{3 \big(4 N^2+6 N-3\big)}{(N+1) (N+2) (2 N+1)}
\sum_{i=1}^N (-2)^{i}\binom{2i}{i} S_{1,2}\big(\frac{1}{2},1,i\big)
%---
\nonumber\\ &&
-\frac{9 \big(4 N^2+6 N-3\big)}{(N+1) (N+2) (2 N+1)}
\sum_{i=1}^N (-2)^{i} \binom{2i}{i} 
S_{1,2}\big(\frac{1}{2},-1,i\big)
%---
+\frac{(N+1)}{(N+2)(2 N+1)}\Biggl[
\nonumber\\ &&
-\sum_{i=1}^N \frac{\displaystyle (-1)^{i} \binom{2i}{i}}{i^3}
- 2 \sum_{i=1}^N \frac{\ds \binom{2i}{i} 
S_1\big(i\big)}{\ds i^2}
+ \frac{3}{2} \sum_{i=1}^N \frac{\ds \binom{2i}{i} S_1^2\big(i\big)}{\ds i}
+ \frac{9}{2} \sum_{i=1}^N \frac{\ds \binom{2i}{i} S_2\big(i\big)}{\ds i}
\nonumber\\ &&
+ 2  \sum_{i=1}^N \frac{\ds (-1)^{i} \binom{2i}{i} S_2\big(i\big)}
{\ds i}
+ 3 \sum_{i=1}^N \frac{\ds \binom{2i}{i} S_{-2}\big(i\big)}{\ds i}
+ 6 \sum_{i=1}^N \frac{\ds (-1)^{i} \binom{2i}{i} 
S_{-2}\big(i\big)}{\ds i}\Biggr]
\Biggr\}
\nonumber\\ &&
+(-1)^N \Biggl\{-\frac{8 \big(N^3+6 N^2+11 
N+7\big)}{3 (N+1)^2 (N+2)^2} S_1^3
+\frac{\big(-4 N^3-7 N^2+6 N+10\big)}{(N+1)^3 (N+2)^3} S_1^2
\nonumber\\ &&
+\Biggl[\frac{2 \big(16 N^3+107 N^2+222 N+146\big)}{(N+1)^4 (N+2)^3}
-\frac{12 \big(N^3+6 N^2+11 N+7\big)}{(N+1)^2 (N+2)^2} S_2\Biggr] S_1
\nonumber\\ &&
+4 S_2^2
+6 S_{-2}^2
+10 S_4
+\frac{2 (3 N+2)}{(N+1)^5 (N+2)^2}
-\frac{8 \big(5 N^3+24 N^2+37 N+20\big)}{3 (N+1)^2 (N+2)^2} S_3
\nonumber\\ &&
-8 (5 N+12) S_5 +8 S_{-4}-10 (N+2) S_{-5}
\nonumber\\ &&
+ \Biggl[-\frac{8 \big(2 N^3+10 N^2+16 N+9\big)}{(N+1)^2 (N+2)^2}
-2 (5 N+12) S_2 - 6 (5 N+12) S_{-2}\Biggr] S_{-3}
\nonumber\\ &&
+ \Biggl[\frac{-36 N^3-165 N^2-270 N-158}{(N+1)^3 (N+2)^3}-2 (5 N+12) S_3-4 S_{2,1}\Biggr] S_2
\nonumber\\ &&
+\frac{4 \big(N^3+6 N^2+11 N+7\big)}{(N+1)^2 (N+2)^2} S_{2,1}
+2 (5 N+12) S_{2,3} + 2(5N+16) S_{2,-3}
-12 S_{3,1} 
\nonumber\\ &&
+16 (N+2) S_{4,1} 
+ \frac{16 \big(N^3+6 N^2+11 N+7\big)}{(N+1)^2 (N+2)^2} S_{-2,1}
+ \Biggl[-\frac{2 \big(4 N^3+7 N^2-6 N-10\big)}{(N+1)^3 (N+2)^3} 
\nonumber\\ &&
-\frac{16 \big(N^3+6 N^2+11 N+7\big)}{(N+1)^2 (N+2)^2} S_1
+4 S_2 +6 (N+4) S_3 +8 (N+2) S_{-2,1} \Biggr] S_{-2}
\nonumber\\ &&
+2 N S_{-2,3} +2 (23 N+60) S_{-2,-3} +4 S_{2,1,1} -16 S_{2,1,-2} +8 S_{2,2,1}
\nonumber\\ &&
+6 (N+4) S_{3,1,1} -8 (N+2) S_{-2,1,-2} -16 S_{2,1,1,1} 
\nonumber\\ &&
 - 2 (3 N+8) \left[S_{1,2} \left(\frac{1}{2},1\right)
               + 3 S_{1,2} \left(\frac{1}{2},-1\right)
\right] S_2(-2)
\nonumber\\ && + 2(3N+8) \Biggl[
- 3 S_{1,4}\left(\frac{1}{2},2\right)
-   S_{1,4}\left(\frac{1}{2},-2\right)
+   S_{1,2,2}\left(\frac{1}{2},1,-2\right)
\nonumber\\ &&
+ 3 S_{1,2,2}\left(\frac{1}{2},-1,-2\right)
+   S_{1,2,2}\left(\frac{1}{2},-2,1\right)
+ 3 S_{1,2,2}\left(\frac{1}{2},-2,-1\right)\Biggr]
\nonumber\\ &&
-6 (3 N+8) \sum_{i=1}^N \ds (-2)^{i} \binom{2i}{i} 
           \left[ S_{1,2}\left(\frac{1}{2},1,i\right)
+3 S_{1,2}\left(\frac{1}{2},-1,i\right)\right]
           \sum_{j=1}^{i} \frac{1}{\ds \binom{2j}{j} 
           j^2}
%---
\nonumber\\ && 
+ 36 \sum_{i=1}^N (-2)^{i} \binom{2i}{i} 
S_{1,2}\left(\frac{1}{2},1,i\right)
\sum_{j=1}^{i} \frac{1}{\ds \binom{2j}{j} j}
%---
\nonumber\\ && 
+108 \sum_{i=1}^N (-2)^{i} \binom{2i}{i} 
S_{1,2}\left(\frac{1}{2},-1,i\right)
\sum_{j=1}^{i} \frac{1}{\ds \binom{2j}{j} j} 
\nonumber\\ && 
%---
+6 (3 N+8)\left[
\sum_{i=1}^N (-2)^{i} \binom{2i}{i} S_{1,2}\left(\frac{1}{2},1,i\right)
%---
+3 
\sum_{i}^N (-2)^{i} \binom{2i}{i} S_{1,2}\left(\frac{1}{2},-1,i\right)\right] 
\sum_{i=1}^N \frac{1}{\ds \binom{2i}{i} i^2}
\nonumber\\ && 
%---
-36 \left[\sum_{i=1}^N (-2)^{i} \binom{2i}{i} S_{1,2}\left(\frac{1}{2},1,i\right)
+ 3 \sum_{i=1}^N (-2)^{i} \binom{2i}{i} S_{1,2}\left(
\frac{1}{2},-1,i\right)\right] 
\sum_{i=1}^N 
\frac{1}{\ds \binom{2i}{i} i}
\nonumber\\ &&
%---
+\frac{3}{2} (3 N+8) \left[ 
\sum_{i=1}^N \frac{
\ds \sum_{j=1}^{i} 
\frac{\ds \binom{2j}{j} S_1^2\big(j\big)}{\ds j}
}
{\ds \binom{2i}{i} \big(1+i\big)}
%---
+ 3 \sum_{i}^N \frac{\ds
\sum_{j=1}^{i} 
\frac{\ds \binom{2j}{j} S_2\big(j\big)}{\ds j}}{\ds
\binom{2i}{i} \big(1+i\big)} \right]
\nonumber\\ &&
%---
+2 (3 N+8) \left[\sum_{i=1}^N \frac{\ds \sum_{j=1}^{i} 
\frac{\ds (-1)^{j} \binom{2j}{j} S_2\big(j\big)}
{\ds j}}{\ds \binom{2i}{i} \big(1+i\big)}
%---
+\frac{3}{2} \sum_{i=1}^N \frac{\ds
\sum_{j=1}^{i} 
\frac{\ds \binom{2j}{j} S_{-2}\big(j\big)}{j}}
{\ds \binom{2i}{i} \big(1+i\big)} \right]
\nonumber\\ &&
+6 (3 N+8) \sum_{i=1}^N \frac{\ds
\sum_{j=1}^{i} 
\frac{\ds (-1)^{j} \binom{2j}{j} S_{-2}\big(j\big)}{\ds j}}
{\ds \binom{2i}{i} \big(1+i\big)}
%---
+2 (3 N+5) \sum_{i=1}^N \frac{\ds \sum_{j=1}^{i} 
\frac{\ds (-1)^{j} \binom{2j}{j}}{j^3}}{\ds
\binom{2i}{i} \big(1+2i\big)}
%---
\nonumber\\ &&
+4 (3 N+5) \sum_{i=1}^N \frac{\ds \sum_{j=1}^{i} \frac{\ds \binom{2j}{j}
S_1\big(j\big)}{j^2}}{\ds \binom{2i}{i} \big(1+2i\big)}
%---
-3 (3 N+5) \sum_{i=1}^N \frac{\ds \sum_{j=1}^{i} 
\frac{\ds \binom{2j}{j} S_1^2\big(j\big)}{\ds j}}{\ds \binom{2i}{i}
\big(1+2i\big)}
\nonumber\\ &&
%---
-9 (3 N+5) \sum_{i=1}^N \frac{\ds \sum_{j=1}^{i} 
\frac{\ds \binom{2j}{j} S_2\big(j\big)}{\ds j}}
{\ds \binom{2i}{i} \big(1+2i\big)}
%---
-4 (3 N+5) \sum_{i=1}^N \frac{\ds \sum_{j=1}^{i} 
\frac{\ds (-1)^{j} \binom{2j}{j} S_2\big(j\big)}
{\ds j}}{\ds \binom{2i}{i} \big(1+2i\big)}
%---
\nonumber\\ &&
-6 (3 N+5) \sum_{i=1}^N \frac{\ds \sum_{j=1}^{i} 
\frac{\ds \binom{2j}{j} S_{-2}\big(j\big)}{\ds j}}{\ds \binom{2i}{i} \big(1+2 i\big)}
%---
-12 (3 N+5) \sum_{i=1}^N \frac{\ds \sum_{j=1}^{i} 
\frac{\ds (-1)^{j} \binom{2j}{j} S_{-2}\big(j\big)}
{\ds j}}{\ds \binom{2i}{i} \big(1+2i\big)}
\nonumber\\ &&
+(-3 N-8) 
\sum_{i=1}^N \frac{\ds
\sum_{j=1}^{i} 
\frac{\ds 
(-1)^{j}\binom{2j}{j}
}{\ds j^3}}
{\ds \binom{2i}{i} \big(1+i\big)}
-2 (3 N+8) 
\sum_{i=1}^N \frac{\ds
\sum_{j=1}^{i} \frac{\ds \binom{2j}{\ds j} 
S_1\big(j\big)}{\ds j^2}}
{\ds \binom{2i}{i} \big(1+i\big)}~~\Biggr\}
\nonumber\\ &&
+\Biggl\{
(-1)^N \Biggl[
6 \big(N^2+1\big) 
\frac{1}{\ds (N-1) N^2 \binom{2 N}{N}} 
\sum_{i=1}^N \ds (-2)^i \binom{2 i}{i}
-6 (3 N-1) \sum_{i=1}^N \frac{\sum_{j=1}^i (-2)^j \ds \binom{2 j}{j}}{\ds i^2 \binom{2 i}{i}}
\nonumber\\ &&
+36 \sum_{i=1}^N \frac{\sum_{j=1}^i \ds (-2)^j \binom{2 j}{j}}{\ds i \binom{2 i}{i}} 
-36 S_1(-2)+8 (3 N-1) S_2(-2)
+\frac{4 \big(N^2-N+1\big)}{(N-1) N^2}
+4 S_2
\nonumber\\ &&
-4 (2 N-1) S_{-2}     
\Biggr]
+\frac{4 \big(N^2-3 N+1\big)}{(N-1) N^2}
-\frac{2^{N+3} \big(N^2-N+1\big)}{(N-1) N^2}
\Biggr\} \zeta_3~.
\label{eqI5b2}
\end{eqnarray}
%---------------------------------------------------------------------------
The integral $I_{7b}(N)$, beyond the harmonic \cite{Vermaseren:1998uu,Blumlein:1998if} and generalized 
harmonic sums \cite{Moch:2001zr,Ablinger:2013cf}
also contains a series of finite binomially and inverse-binomially nested sums, summing over
generalized harmonic sums. These structures emerge from the hyperlogarithms containing the
set of letters in the alphabet (\ref{eq:alphab}) beyond those of harmonic polylogaritms and the
root-function $r(x)$ in the argument. It is the strength of packages like {\tt Sigma} \cite{SIGMA}
based on general summation algorithms operating on difference fields to find the new sum-structures.
Furthermore, the representation (\ref{eqI5b2}) is given by sums being transcendental to each other. Here we made 
use of sum representations
having been introduced previously in 
Refs.~\cite{Vermaseren:1998uu,Blumlein:1998if,Ablinger:2011te,Moch:2001zr,Ablinger:2013cf}
which occur at lower levels of the sum hierarchy 
implied by Feynman integrals. 
%%%%%%%%%%%%%%%%%%%%%%%%%%%%%%%%%%%%%%%%%%%%%%%%%%%%%%%%%%%%%%%%%%%%%%%
\section{Analytic Continuation of Binomially Weighted Nested Sums}
\label{sec:an}
\renewcommand{\theequation}{\thesection.\arabic{equation}}
\setcounter{equation}{0} 
%%%%%%%%%%%%%%%%%%%%%%%%%%%%%%%%%%%%%%%%%%%%%%%%%%%%%%%%%%%%%%%%%%%%%%%

\vspace{1mm} \noindent
To obtain the analytic continuation of the binomial sums as given in (\ref{eqI5b2}) we first
derive their representation in terms of a Mellin transformation
%---------------------------------------------------------------------------
\begin{eqnarray}
\Mvec[f(x)](N) = \int_0^1 dx x^N~f(x)~.
\end{eqnarray}
%---------------------------------------------------------------------------
Individual nested sums then usually are given as a linear combination
%---------------------------------------------------------------------------
\begin{equation}\label{eq:IntRepForm}
c_0 + \sum_{j=1}^k c_j^N \Mvec[f_j(x)](N),
\end{equation}
%---------------------------------------------------------------------------
where the constants $c_j$ and functions $f_j(x)$ do not depend on $N$. The functions $f_j(x)$
are defined in terms of iterated integrals. 
As starting point we 
only need the following basic integral representations~:
%----------------------------------------------------------------------------
\begin{eqnarray}
\label{eq:MEL1}
\frac{1}{N} &=& \Mvec\left[\frac{1}{x}\right](N)
\\
%-------
\binom{2N}{N} &=& \frac{4^N}{\pi}\Mvec[f_{\sf w_1}(x)](N)
\\
%-------
\label{eq:MEL2}
\frac{1}{N\displaystyle\binom{2N}{N}} &=& \frac{1}{4^N}\Mvec[f_{\sf w_3}(x)](N),
\end{eqnarray}
%----------------------------------------------------------------------------
where the letters $f_{\sf w_i}(x)$ are given by
%----------------------------------------------------------------------------
\begin{eqnarray}
f_{\sf w_1}(x) &=& \frac{1}{\sqrt{x} \sqrt{1-x}} \\
f_{\sf w_3}(x) &=& \frac{1}{x \sqrt{1-x}}~.
\end{eqnarray}
%----------------------------------------------------------------------------
Here and in the following we refer to the notation of Ref.~\cite{BINOM}.

From the Mellin transforms (\ref{eq:MEL1}--\ref{eq:MEL2})
we can obtain integral representations for the nested sums step by step. 
In general the computation proceeds as follows. Starting from the innermost sum we move 
outwards maintaining an integral representation of the sub-expressions visited so far. 
For each intermediate sum
%----------------------------------------------------------------------------
\begin{equation}
\label{eq:AlgorithmIntermediate}
\sum_{i_j=1}^Na_j(i_j)\sum_{i_{j+1}=1}^{i_j} a_{j+1}(i_{j+1})\dots\sum_{i_k=1}^{i_{k-1}}a_k(i_k)
\end{equation}
%----------------------------------------------------------------------------
this first involves setting up an integral representation for the summand $a_j(N)$ of the 
form \eqref{eq:IntRepForm}. This may require computation of Mellin convolutions, which we 
will describe in more detail below. Next we obtain an integral representation of the same 
form of
%----------------------------------------------------------------------------
\begin{equation}
a_j(N)\sum_{i_{j+1}=1}^Na_{j+1}(i_{j+1})\dots\sum_{i_k=1}^{i_{k-1}}a_k(i_k)
\end{equation}
%----------------------------------------------------------------------------
by Mellin convolution with the result for the inner sums computed so far. Then by the 
summation property 
%----------------------------------------------------------------------------
\begin{equation}
\sum_{i=1}^N c^i \Mvec[f(x)](i) = c^N \Mvec\left[\frac{x}{x-\tfrac{1}{c}} f(x) \right](N) 
-\Mvec\left[\frac{x}{x-\tfrac{1}{c}} f(x) \right](0)
\end{equation}
%----------------------------------------------------------------------------
we obtain an integral representation for 
the sum \eqref{eq:AlgorithmIntermediate}. These steps are repeated until the outermost 
sum has been processed.

Now, we take a closer look at how we compute Mellin convolutions, which is the most 
challenging part of the calculation. Formally, we rely on the convolution formulae
%----------------------------------------------------------------------------
\begin{eqnarray}
\Mvec[f(x) * g(x)](N) &=& 
\Mvec[f(x)](N) \Mvec[g(x)](N)\\
f(x) * g(x) &=& \int_0^1 dx_1 \int_0^1 dx_2 \delta(x - x_1 x_2) f(x_1) g(x_2),  
\end{eqnarray}
%----------------------------------------------------------------------------
which give us a definite integral depending on a continuous 
parameter and which can be written in the form
%----------------------------------------------------------------------------
\[
F(x)=\int_x^1dyf(x,y).
\]
%----------------------------------------------------------------------------
In order to obtain a closed form for this integral, we first set up a differential 
equation satisfied by $F(x)$ and then obtain a solution of this equation satisfying 
appropriate initial conditions. In the first step we exploit the principle of 
differentiation under the integral. If we have a relation for the integrand $f(x,y)$ 
of the form
%----------------------------------------------------------------------------
\begin{equation}\label{eq:ParametricIntegration}
c_m(x)\frac{\partial^mf}{\partial{x}^m}(x,y)+\dots+c_0(x)f(x,y)
=\frac{\partial g}{\partial{y}}(x,y)
\end{equation}
%----------------------------------------------------------------------------
for some coefficients $c_i(x)$ independent of $y$ and some function $g(x,y)$, then 
by applying the operator $\int_x^1dy$ this gives rise to a linear ordinary differential equation 
for the integral $F(x)$
%----------------------------------------------------------------------------
\begin{equation}\label{eq:ResultingODE}
c_m(x)F^{(m)}(x)+\dots+c_0(x)F(x)=g(x,1)-g(x,x)+\text{additional boundary terms}.
\end{equation}
%----------------------------------------------------------------------------
Proper care has to be taken for evaluating the right hand side of this relation in 
the presence of singularities. There are several computer algebra algorithms for 
different types of integrands $f(x,y)$ which, given $f(x,y)$, compute relations 
of the form \eqref{eq:ParametricIntegration}. They either utilize differential 
fields \cite{Risch,Bronstein,RaabPhD} or holonomic systems and Ore algebras 
\cite{AlmkvistZeilberger,Chyzak,Koutschan}. For obtaining solutions to the generated 
differential equations the following two observations are crucial. All differential 
equations obtained during our computations factor completely into first-order equations 
with rational function coefficients and, moreover, these factors all have algebraic 
functions of degree at most two as their solutions. These two observations imply 
that solutions are of the form
%----------------------------------------------------------------------------
\[
\frac{r_1(x)}{\sqrt{p_1(x)}} \int dx \frac{r_2(x)}{\sqrt{p_2(x)}} 
\int dx \dots \int dx \frac{r_k(x)}{\sqrt{p_k(x)}},
\]
%----------------------------------------------------------------------------
where $r_i(x)$ are rational functions and $p_i(x)$ are square-free polynomials. 
We define the iterated integrals ${\rm H}^*_{\vec{\sf w}}(x)$ by
%----------------------------------------------------------------------------
\begin{eqnarray}
{\rm H}^*_{{\sf a}, {\vec{\sf b}}}(x) = \int_x^1 dy f_{\sf a}(y) {\rm H}^*_{\vec{\sf b}}(y),~~~~~~
{\rm H}^*_\emptyset(x) = 1~, 
\end{eqnarray}
%----------------------------------------------------------------------------
and $f_{\sf j}(x)$ are the corresponding basic functions, which partly contain 
root-valued denominators. 
Using a dedicated rewrite procedure \cite{HERMITE} based on integration by parts we can write 
a basis of the solution space in terms of the iterated integrals 
which is then used to match initial conditions.

For the representation of integral $I_{7b}$ 32 different letters $f_{\sf j}(x)$ are 
needed, cf. \cite{BINOM}.
As an example we consider the representation for the following double sum~:
%----------------------------------------------------------------------------
\begin{eqnarray}
\lefteqn{\sum_{i=1}^N \frac{(-1)^i}{\ds (2i+1) \binom{2i}{i}} \sum_{j=1}^i \binom{2j}{j} \frac{S_2(j)}{j} 
 =} \nonumber\\ && \frac{1}{2} (-1)^N 
\Mvec\left[\frac{x(-{\rm H}^*_{\sf w_8,1,0}(x) + \zeta_2 {\rm H}^*_{\sf 
w_8}(x))}{(x+1)\sqrt{x-\tfrac{1}{4}}}\right](N)
- \frac{1}{2}\Mvec\left[\frac{x(-{\rm H}^*_{\sf w_8,1,0}(x) + \zeta_2 {\rm H}^*_{\sf 
           w_8}(x))}{(x+1)\sqrt{x-\tfrac{1}{4}}}\right](0)
\nonumber\\ &&
- \frac{\zeta_3}{3} \left(-\frac{1}{4}\right)^N \Mvec\left[\frac{x}{(x+4)\sqrt{1-x}}\right](N)
+ \frac{\zeta_3}{3} \Mvec\left[\frac{x}{(x+4)\sqrt{1-x}}\right](0)~.
\end{eqnarray}
%----------------------------------------------------------------------------
Here the last Mellin-transform at argument $N=0$ takes the value 
$2 + (8/\sqrt{5})[\ln(\sqrt{5}-1) - \ln(2)]$, while the former one is a new constant, beyond 
the (cyclotomic) multiple zeta values. The letter $f_{\sf w_8}$ is given by
%----------------------------------------------------------------------------
\begin{eqnarray}
f_{\sf w_8}(x) = \frac{1}{x \sqrt{x - \tfrac{1}{4}}}~.
\end{eqnarray}
%----------------------------------------------------------------------------
To perform the asymptotic expansion of the Mellin-transforms we use the representation
%----------------------------------------------------------------------------
\begin{eqnarray}
\label{eq:MELAS}
\Mvec[f(x)](N) = \int_0^\infty dz e^{-Nz} f\left(e^{-z}\right) e^{-z}.
\end{eqnarray}
%----------------------------------------------------------------------------
One may expand $f\left(e^{-z}\right) e^{-z}$ at $z=0$ and integrate (\ref{eq:MELAS}) term-wise
to obtain the asymptotic expansion for $|N| \rightarrow \infty, {\rm arg}(N) \neq \pi$ using
%----------------------------------------------------------------------------
\begin{eqnarray}
\int_0^\infty dz e^{-Nz} z^c \ln^k(z) = \frac{\partial^k}{\partial c^k} \frac{\Gamma(c+1)}{N^{c+1}}
= \sum_{i=0}^k (-1)^i \binom{k}{i} \Gamma^{(k-i)}(c+1) \frac{\ln^i(N)}{N^{c+1}}
\end{eqnarray}
%----------------------------------------------------------------------------
for $c > -1$ and $k \in \mathbb{N}$. These expansions are automated in the package {\tt HarmonicSums}.
With these prerequisites at hand the asymptotic expansion of (\ref{eqI5b2}) can now be performed.

It turns out, that part of the individual sums  contributing to (\ref{eqI5b2}) diverge
$\propto 8^N, 4^N$ and $2^N$ for large values of $N$. In case of the present scalar integral 
$I_{7b}(N)$ the terms $\propto 8^N$ and $\propto 4^N$ cancel, while some of the terms $\propto 2^N$ 
remain. We also have checked the principal divergence of this graph for $N \rightarrow \infty$ numerically. 
In the 
physical case, accounting for all color and numerator structures, also these terms 
are expected to cancel between the different diagrams. Due to the contributing large class of new sums 
one expects also that a series of new constants beyond the multiple zeta values \cite{Blumlein:2009cf}, 
generalized (cyclotomic) zeta values \cite{Ablinger:2011te,Ablinger:2013cf} contribute, see also \cite{BINOM}. 

The asymptotic representation of $I_{7b}(N)$ reads~:
%---------------------------------------------------------------------------
\begin{eqnarray}
I_{7b}(N) \propto 2^N \hat{I}_{7b,1}(N) + \hat{I}_{7b,2}(N)~, 
\label{eqI5bc}
\end{eqnarray}
%---------------------------------------------------------------------------
with
%---------------------------------------------------------------------------
\begin{eqnarray}
{I}_{7b,1}(N) &\simeq&
\Biggl[-\frac{112}{9 N^3}
+\frac{7568}{81 N^4}
-\frac{27280}{81 N^5}
+\frac{2256112}{2187 N^6}
-\frac{52719920}{19683 N^7}
+\frac{373195088}{59049 N^8}\Biggr] \zeta_3
\\
%-------------------------------------------------------------------------------------
{I}_{7b,2}(N) &\simeq&
\Biggl[
\frac{1}{N^4}
-\frac{12}{N^5}
+\frac{91}{N^6}
-\frac{574}{N^7}
+\frac{3451}{N^8}
\Biggr] \zeta_2
\nonumber\\ &&
+2^{-N} 
\Biggl[
\Biggr[
-\frac{3}{2 N^2}
+\frac{1}{2 N^3}
+\frac{6}{N^4}
-\frac{35}{2 N^5}
+\frac{17}{N^6}
+\frac{79}{2 N^7}
-\frac{152}{N^8}
\Biggr] \ln^2(2) 
\nonumber\\ &&
+\Biggl[
-\frac{3}{N^2}
+\frac{1}{N^3}
+\frac{12}{N^4}
-\frac{35}{N^5}
+\frac{34}{N^6}
+\frac{79}{N^7}
-\frac{304}{N^8}
\Biggr] \left( \Li_2\left(-\frac{1}{2}\right)
+\frac{1}{2} \zeta_2\right)
\nonumber\\ &&
+\Biggl[
-\frac{3}{2 N^2}
+\frac{1}{2 N^3}
+\frac{6}{N^4}
-\frac{35}{2 N^5}
+\frac{17}{N^6}
+\frac{79}{2 N^7}
-\frac{152}{N^8}
\Biggr] 
\zeta_2
\Biggr]
+
\Biggl[
 \frac{2}{N^2}
-\frac{6}{N^3}
+\frac{8}{N^4}
+\frac{14}{N^5}
\nonumber\\ &&
-\frac{128}{N^6}
+\frac{478}{N^7}
-\frac{1272}{N^8}
\Biggr]
\zeta_3
+(-1)^N 
\Biggl[
\Biggl[
-\frac{4}{3 N^2}
+\frac{52}{9 N^3}
-\frac{56}{3 N^4}
+\frac{2396}{45 N^5}
-\frac{424}{3 N^6}
+\frac{22516}{63 N^7}
\nonumber\\ &&
-\frac{872}{N^8}
\Biggr]
\ln^3(\bar{N})
+\Biggl[
-\frac{74}{9 N^3}
+\frac{133}{3 N^4}
-\frac{4103}{25 N^5}
+\frac{15439}{30 N^6}
-\frac{6456953}{4410 N^7}
+\frac{1230668}{315 N^8}
\Biggr]
\ln^2(\bar{N})
\nonumber\\ &&
+
\Biggl[
\Biggl[
-\frac{2}{N^2}
+\frac{26}{3 N^3}
-\frac{28}{N^4}
+\frac{1198}{15 N^5}
-\frac{212}{N^6}
+\frac{11258}{21 N^7}
-\frac{1308}{N^8}
\Biggr] 
\zeta_2
+\frac{4}{N^2}
-\frac{436}{27 N^3}
+\frac{29}{N^4}
\nonumber\\ &&
+\frac{32}{375 N^5}
-\frac{8489}{36 N^6}
+\frac{8193131}{6860 N^7}
-\frac{778753}{180 N^8}
\Biggr]
\ln(\bar{N})
+ A_1 + A_2 N
+\Biggl[
-\frac{8}{N}
+\frac{21}{N^2}
-\frac{520}{9 N^3}
\nonumber\\ &&
+\frac{476}{3 N^4}
-\frac{21473}{50 N^5}
+\frac{68569}{60 N^6}
-\frac{26328833}{8820 N^7}
+\frac{4823873}{630 N^8}
\Biggr]
\zeta_2
+2^{-N} 
\Biggl[
\Biggl[
\Biggl[
-\frac{3}{2}
-\frac{1}{N}
-\frac{1}{N^2}
\nonumber\\ &&
+\frac{15}{N^3}
-\frac{121}{N^4}
+\frac{1023}{N^5}
-\frac{9721}{N^6}
+\frac{104415}{N^7}
-\frac{1259161}{N^8}
\Biggr]
\zeta_2
-\frac{3}{N}
+\frac{11}{2 N^2}
-\frac{55}{2 N^3}
+\frac{602}{3 N^4}
\nonumber\\ &&
-\frac{50497}{30 N^5}
+\frac{239851}{15 N^6}
-\frac{36068621}{210 N^7}
+\frac{43495976}{21 N^8}
\Biggr]
\ln^2(2)
+\Biggl[
-\frac{3}{2}
-\frac{1}{N}
-\frac{1}{N^2}
+\frac{15}{N^3}
\nonumber\\ &&
-\frac{121}{N^4}
+\frac{1023}{N^5}
-\frac{9721}{N^6}
+\frac{104415}{N^7}
-\frac{1259161}{N^8}
\Biggr]
\zeta_2^2
+\Biggl[
-\frac{3}{N}
+\frac{11}{2 N^2}
-\frac{55}{2 N^3}
+\frac{602}{3 N^4}
\nonumber\\ &&
-\frac{50497}{30 N^5}
+\frac{239851}{15 N^6}
-\frac{36068621}{210 N^7}
+\frac{43495976}{21 N^8}
\Biggr]
\zeta_2
+\left(
\Li_2\left(-\frac{1}{2}\right) + \frac{1}{2} \zeta_2\right)
\nonumber\\ &&
\times
\Biggl[
\Biggl[
-3
-\frac{2}{N}
-\frac{2}{N^2}
+\frac{30}{N^3}
-\frac{242}{N^4}
+\frac{2046}{N^5}
-\frac{19442}{N^6}
+\frac{208830}{N^7}
-\frac{2518322}{N^8}
\Biggr] 
\zeta_2
\nonumber\\ &&
-\frac{6}{N}
+\frac{11}{N^2}
-\frac{55}{N^3}
+\frac{1204}{3 N^4}
-\frac{50497}{15 N^5}
+\frac{479702}{15 N^6}
-\frac{36068621}{105 N^7}
+\frac{86991952}{21 N^8}
\Biggr]
\Biggr]
\nonumber\\ &&
+\Biggl[
-\frac{2}{N}
+\frac{10}{3 N^2}
-\frac{46}{9 N^3}
+\frac{20}{3 N^4}
-\frac{242}{45 N^5}
-\frac{20}{3 N^6}
+\frac{3194}{63 N^7}
-\frac{180}{N^8}
\Biggr]
\zeta_3
\nonumber\\ &&
+\frac{6}{N^2}
-\frac{1732}{81 N^3}
+\frac{793}{12 N^4}
-\frac{1217029}{5625 N^5}
+\frac{130343}{180 N^6}
-\frac{10153834441}{4321800 N^7}
+\frac{1632850801}{226800 N^8}
\Biggr]
\nonumber\\ &&
+\frac{4}{N^5}
-\frac{62}{N^6}
+\frac{1759}{3 N^7}
-\frac{4530}{N^8}
\nonumber\\ &&
+
\Biggl\{
-\frac{2}{N^2}
-\frac{6}{N^3}
-\frac{8}{N^4}
-\frac{2}{N^5}
+\frac{8}{N^6}
-\frac{10}{N^7}
-\frac{72}{N^8}
\nonumber\\ &&
+(-1)^N \Biggl[
\frac{10}{3} \zeta_2
-\frac{4 \pi }{\sqrt{3}}
+\frac{2}{N^2}
+\frac{10}{3 N^3}
+\frac{4}{N^4}
+\frac{62}{15 N^5}
+\frac{4}{N^6}
+\frac{82}{21 N^7}
+\frac{4}{N^8}
\Biggr] 
\nonumber\\ &&
+\left(-\frac{1}{4}\right)^N \sqrt{\pi } 
\Biggl[
-\frac{64}{3} \left(\frac{1}{N}\right)^{5/2}
+\frac{232}{9} \left(\frac{1}{N}\right)^{7/2}
-\frac{6697}{54} \left(\frac{1}{N}\right)^{9/2}
+\frac{65167}{144} \left(\frac{1}{N}\right)^{11/2}
\nonumber\\ &&
-\frac{30311555}{13824} \left(\frac{1}{N}\right)^{13/2}
+\frac{3942221963}{331776}\left(\frac{1}{N}\right)^{15/2}
\Biggr]\Biggr\} \zeta_3~.
\label{eqI5bd}
\end{eqnarray}
%---------------------------------------------------------------------------
Here the constants $A_1$ and $A_2$ are given by
%---------------------------------------------------------------------------
\begin{eqnarray}
A_1 &=& 
18.6886524505148659%00722420496284579915565  
+  16 H_{-1, 0, 2, 1, 0}(1) 
+ 48 \Biggl[H_{0, -2, -1, 0, 1}(1) 
+  H_{0, -2, -1, 1, 0}(1) 
\nonumber\\ && +  H_{0, -2, 1, -1, 0}(1) 
+  H_{0, 1, -2, -1, 0}(1)\Biggr] 
\\
A_2 &=& 
4.67069037753751178 %4948178677388905554018 
+ 6  H_{-1, 0, 2, 1, 0}(1) + 
 18 [H_{0, -2, -1, 0, 1}(1) + H_{0, -2, -1, 1, 0}(1) 
\nonumber\\ &&
 + 
 H_{0, -2, 1, -1, 0}(1) +  H_{0, 1, -2, -1, 0}(1)]~.
\end{eqnarray}
%---------------------------------------------------------------------------
We have expressed part of these constants numerically up to five generalized harmonic
polylogarithms at $x=1$.
We checked using PSLQ \cite{PSLQ} that no integer relation between these HPLs based on 100 digits 
is found. The numerical values of these constants can be derived from the following
one-dimensional integral representations referring to classical polylogarithms
%---------------------------------------------------------------------------
\begin{eqnarray}
H_{-1, 0, 2, 1, 0}(1) &=& 
\int_0^1 dx \frac{(\Li_2(1-x)- \zeta_2)(\Li_2(-x)+\log(2) \log(x)+\zeta_2/2)}{x-2}
\nonumber\\
&=&-0.07640650747463134675 %
%36238843473363326930153844361007331968301730780\
%77090829433442765736923525443241104259611611
\\
H_{0, -2, -1, 0, 1}(1) &=& \int_0^1 dx
\frac{\Li_2(x)[\Li_2(-x/2)-\Li_2(-1/2)-\log(x) \log(2/(x+2))]}{x+1}
\nonumber\\ &=&
    0.01812205214208962744
%6780495495576352385571310741982902559901\
%3277261041483488922389572858799615918806472711166615923692461799770204\
%64105
\\
H_{0, -2, -1, 1, 0}(1) &=& \int_0^1 dx  
\frac{(\Li_2(1-x)- \zeta_2)[\Li_2(-x/2)-\Li_2(-1/2)-\log(x) \log(2/(x+2))]}{x+1}
\nonumber\\ &=&
   -0.04281095672416394220
%344829295703833844699412330302783200486\
%6552903537976279226684917860286114703872082548000505774328745465728550\
%223081
\\
H_{0, -2, 1, -1, 0}(1) &=& 
\int_0^1 dx [\Li_2(-x) + \ln(x) \ln(1+x)] \nonumber\\ && \times
\frac{\Li_2(-x/2)-\Li_2(-1/2)-\log(x) \log(2/(2+x))}{1-x}
\nonumber\\ 
&=& -0.07000199841995163532
%280287504084337369760800646050004659573\
%2183937537705761626327189876162492770795583986697363283325449647911828\
%247417
\\
H_{0, 1, -2, -1, 0}(1) 
&=&
\int_0^1 dx \frac{\Li_2(1-x) [\Li_2(-x)+\log(x)\log(x+1)]}{x+2}
\nonumber\\ &=&
    -0.13932305992518092238~.
%866462590352172678872199187023412800167\
%3085595651514094711693948594380866603996325847505753139564084240283109\
%255005
\end{eqnarray}
%---------------------------------------------------------------------------
The numerical parts recruit from 20 one- and 17 two-dimensional  integrals,
which will be  given in \cite{BINOM} in explicit form. One example
reads
%---------------------------------------------------------------------------
\begin{eqnarray}
 \lefteqn{\Mvec\left[\frac{x\HA_{\sf w_8,0,1}(x)}{(x-1)\sqrt{x-\frac{1}{4}}}\right](0)\ =\ \int_{\frac{1}{4}}^1dx\frac{\left(\sqrt{4x-1}-\frac{2}{\sqrt{3}}\arccosh\left(\frac{2x+1}{2(1-x)}\right)\right)\left(\text{Li}_2(x)-\zeta_2\right)}{x\sqrt{x-\frac{1}{4}}}}\nonumber\\
 &&+4\int_0^1dx\frac{\left(\sqrt{x(2-x)}-1-\frac{2}{\sqrt{3}}\left(\arccos\left(\frac{x^2-2x+3}{(3-x)(x+1)}\right)-\frac{\pi}{3}\right)\right)\left(\text{Li}_2\left(\frac{(1-x)^2}{4}\right)-\zeta_2\right)}
{(1-x)\sqrt{x(2-x)}}~.
\end{eqnarray}
%---------------------------------------------------------------------------
Furthermore, beyond the usual multiple zeta values \cite{Blumlein:2009cf}, also the following
constants contribute
%---------------------------------------------------------------------------
\begin{eqnarray}
&& \Biggl\{
\sqrt{3}, 
\sqrt{\pi}, 
\ln(\sqrt{5}-1),
\ln(2 - \sqrt{3}),
\ln(3), 
\Li_2\left(-\frac{1}{2}\right), 
\Li_3\left(-\frac{1}{3}\right), 
\Li_3\left(-\frac{1}{2}\right), 
\Li_3\left(\frac{3}{4}\right), 
\nonumber\\ &&
\Li_3\left(\frac{\sqrt{5}-1}{2}\right),
\Li_4\left(\frac{1}{4}\right), 
\psi'\left(\frac{1}{3}\right)
\Biggr\}~.
\end{eqnarray}
%---------------------------------------------------------------------------
Some of the latter constants express infinite cyclotomic harmonic 
sums \cite{Ablinger:2011te} or represent the infinite sums
%---------------------------------------------------------------------------
\begin{eqnarray}
 S_{1,1}\left(\frac{1}{2},\frac{1}{2};\infty\right) &=& \frac{1}{2} 
\zeta_2 +\text{Li}_2(-\tfrac{1}{2})
\\
 S_{1,2}\left(-\frac{1}{2},-\frac{1}{2};\infty\right) &=&
\text{Li}_3(\tfrac{3}{4})+\text{Li}_3(\tfrac{2}{3})
+3\text{Li}_3(-\tfrac{1}{2})+\text{Li}_2(-\tfrac{1}{2})[\ln(3)-3\ln(2)]\nonumber\\
 &&+\frac{7}{8}\zeta_3-\frac{3}{2}\zeta_2\ln(2)-\frac{1}{6} \ln^3(3)
-\frac{19}{6}\ln^3(2)-\ln(2)\ln^2(3)
\nonumber\\ &&
+\frac{9}{2}\ln^2(2) \ln(3)~.
\label{eq:Li3n}
\end{eqnarray}
%---------------------------------------------------------------------------
Eq. (\ref{eq:Li3n}) can be further simplified using the relation
%---------------------------------------------------------------------------
\begin{eqnarray}
\Li_3\left(\tfrac{2}{3}\right) &=& 
  \frac{1}{2} \Li_3\left(\tfrac{3}{4}\right)
+ \Li_3\left(-\tfrac{1}{2}\right) 
+ \frac{7}{6} \zeta_3 + \ln(3) \ln^2(2) - \frac{1}{2} \ln(2) \ln^2(3) + \frac{1}{6} \ln^3(3)
- \frac{5}{6} \ln^3(2) 
\nonumber\\ &&
- \zeta_2 [\ln(3) - \ln(2)]
\end{eqnarray}
%---------------------------------------------------------------------------
found first by applying PSLQ \cite{PSLQ} on the basis of 100 digits and checked for $10^4$ digits.
We derived this relation also analytically.
Also the integral 
%---------------------------------------------------------------------------
\begin{eqnarray}
 \mbox{}_5F_4\left({1,1,1,1,\frac{3}{2} \atop 2,2,2,2} \middle| -4 \right) &=&
\frac{1}{4}\int_0^1dt\left(1-\frac{1}{\sqrt{1+4t}}\right)\frac{\ln(t)^2}{t}\nonumber\\
 &=& 6\text{Li}_3\left(\tfrac{\sqrt{5}-1}{2}\right)
+4\text{Li}_3\left(-\tfrac{\sqrt{5}-1}{2}\right)
%\nonumber\\ &&
-2\zeta_3
-2 \zeta_2
\ln\left(\tfrac{\sqrt{5}-1}{2}\right)
+\frac{4}{3}\ln^3\left(\tfrac{\sqrt{5}-1}{2}\right)
\label{eq:GR}
\nonumber\\
\end{eqnarray}
%---------------------------------------------------------------------------
contributes, containing polylogarithms at the inverse of the golden ratio $(\sqrt{5}-1)/2  = 
2/(\sqrt{5}+1)$. One may further simplify (\ref{eq:GR}) using the identity \cite{TRILOG}
%---------------------------------------------------------------------------
\begin{eqnarray}
\Li_3\left[\left(\tfrac{\sqrt{5}-1}{2}\right)^2\right] &=& \frac{4}{5} \zeta_3 +
\frac{2}{3} \ln^3\left(\tfrac{\sqrt{5}+1}{2}\right)
-\frac{2}{15} \pi^2 \ln\left(\tfrac{\sqrt{5}+1}{2}\right)~.
\end{eqnarray}
%---------------------------------------------------------------------------
Furthermore, half-integer powers appear in the asymptotic expansion (\ref{eqI5bd}).
%%%%%%%%%%%%%%%%%%%%%%%%%%%%%%%%%%%%%%%%%%%%%%%%%%%%%%%%%%%%%%%%%%%%%%%
\section{Moments for Crossed-Box Graphs}
\label{sec:c}
\renewcommand{\theequation}{\thesection.\arabic{equation}}
\setcounter{equation}{0} 
%%%%%%%%%%%%%%%%%%%%%%%%%%%%%%%%%%%%%%%%%%%%%%%%%%%%%%%%%%%%%%%%%%%%%%%

\vspace{1mm} \noindent
Using the method of hyperlogarithms also fixed moments of convergent graphs can be evaluated.
The method relies on partial fractioning of the operator polynomial induced by the operator.
Correspondingly, for large values of $N$, the number of terms grows exponentially. The calculation
time and the requested storage are growing significantly. To illustrate the potential of the method 
in this respect we select the possibly most complicated  graphs of the present physics project belonging to 
crossed box topologies. 
%-------------------------------------------------------------------------------------------
%-------------------------------------------------------------------------------------------
\noindent
While for more simple topologies more moments can be calculated
given typical CPU times of various hours to days, in case of the above topologies
the 9th moments could be calculated in about 8 hours requesting a storage of 35 Gbyte. 
The 10th moment would have needed storage of more than 200 Gbyte RAM due to the intense use 
of partial fractioning. Since the algorithm is implemented as {\tt Maple}-code the 
available RAM is the limiting parameter, unlike the case e.g. for {\tt FORM}-programs, 
using also fast external discs \cite{Vermaseren:2000nd}. In comparison, the {\tt FORM}-based 
program {\tt MATAD} \cite{Steinhauser:2000ry} allows to calculate a few higher moments 
as well having the same time- and storage resources.

\renewcommand{\arraystretch}{1.50}
\begin{table}[H]\centering
\begin{tabular}{ |c||c | c | c | }
  \hline                        
  N& (a) & (b) & (c)\\
\hline \hline
  $0$ & 
  {\fs $\frac{1}{4}$}&
  {\fs $\frac{1}{4}$}& 
  {\fs $\frac{1}{4}$}\\ 
\hline
  $1$ &
  {\fs $-\frac{1}{8}$}& 
  {\fs $\frac{1}{16}-\frac{7}{32} \zeta_3$}&
  {\fs $-\frac{3}{16}+\frac{7}{32} \zeta_3$}\\ 
\hline
  $2$ &{\fs $\frac{145}{1536}-\frac{145}{9216} \zeta_3$}& {\fs $-\frac{89}{4608}+\frac{445}{3072} \zeta_3$ 
}&{\fs $-\frac{935}{13824}+\frac{87}{1024} \zeta_3$} \\ \hline
  $3$ &{\fs $-\frac{81}{1024}+\frac{145}{6144} \zeta_3$}&{\fs $\frac{8519}{55296}-\frac{2813}{12288} 
\zeta_3$} &{\fs $-\frac{4993}{36864}+\frac{3145}{24576} \zeta_3$} \\ 
\hline
  $4$ &{\fs $\frac{5582479}{82944000}-\frac{10489}{409600} \zeta_3
$}&{\fs $\frac{369197}{5529600} + \frac{18623}{737280} \zeta_3
$} &{\fs $\frac{2379019}{82944000}-\frac{719}{49152} \zeta_3$} \\ 
\hline
  $5$ &{\fs $	-\frac{1899679}{33177600}
+ \frac{36401}{1474560} \zeta_3
$}&{\fs
    $\frac{18015269}{99532800}
-\frac{4794311}{22118400} \zeta_3
$} &{\fs
    $-\frac{39045971}{298598400}+\frac{507679}{4423680} \zeta_3$} \\ \hline
  $6$ &{\fs $
\frac{141912342181}{2913258700800}
-\frac{695736571}{30828134400} \zeta_3
$}&{\fs
    $\frac{278864978351}{1248539443200}-\frac{5175109523}{39636172800} \zeta_3$} &{\fs
    $\frac{1058933976943}{8739776102400}-\frac{255461723}{2642411520} \zeta_3$} \\ \hline
  $7$ &{\fs $-\frac{11526313783}{277453209600}
+\frac{59076777}{2936012800} \zeta_3
$}&{\fs
    $\frac{25191975655421}{74912366592000}
-\frac{120819716411}{369937612800} \zeta_3$} &{\fs
    $-\frac{7247023939349}{33294385152000}+\frac{370501349}{2013265920} \zeta_3$} \\ \hline
  $8$ &{\fs $\frac{266608033463}{7491236659200}
-\frac{29536680029}{1664719257600} \zeta_3
$}&{\fs 
$\frac{47884345670443}{89894839910400}
-\frac{1916259725321}{4756340736000} \zeta_3
$} &{\fs 
$\frac{13258221091439}{
44947419955200}-\frac{115670928497}{475634073600} \zeta_3$} \\ \hline
  {$9$} &{\fs $-\frac{255303766759}{8323596288000}
+\frac{5768976713}{369937612800} \zeta_3
$} & {\fs
$\frac{49979032484264647}{62926387937280000}
-\frac{75636078173}{108716359680} \zeta_3$}& {\fs
$-\frac{3310967262876383}{6991820881920000}+\frac{4778989541}{12079595520} \zeta_3$}\\
  \hline  
\end{tabular}
\caption[]{\sf Moments of the finite crossed-box graphs (a--c) shown in Figure~9.} 
\end{table}
\renewcommand{\arraystretch}{1.0}
%%%%%%%%%%%%%%%%%%%%%%%%%%%%%%%%%%%%%%%%%%%%%%%%%%%%%%%%%%%%%%%%%%%%%%%
\section{Conclusions}
\label{sec:con}
\renewcommand{\theequation}{\thesection.\arabic{equation}}
\setcounter{equation}{0} 
%%%%%%%%%%%%%%%%%%%%%%%%%%%%%%%%%%%%%%%%%%%%%%%%%%%%%%%%%%%%%%%%%%%%%%%

\vspace*{1mm}
\noindent
It has long been noticed that many results for zero- and single-scale processes in renormalizable
Quantum Field Theories can be expressed in terms of iterated integrals or nested harmonic 
sums at the lower loop level \cite{STRUCT}. Ideally, a direct method was sought for to arrive at these results
right form the Feynman parameterization of the contributing diagrams. In case of convergent Feynman 
integrals the method of hyperlogarithms provides this way in case a Fubini sequence can be found
for the diagram being considered. In the present paper we have extended this method to the case of
massive diagrams including local operator insertions. 

The calculation of fixed moments does not pose a theoretical problem, since the expressions can be reduced in 
principle by applying partial fractioning. With growing values of $N$ the complexity of the expressions rises 
significantly such that the corresponding number of terms cannot be swallowed even by modern computers anymore. 
To extend the present abilities, special software implementations outside coding systems based on {\tt 
Mathematica} and/or {\tt Maple} are necessary, to free the main storage and allow the use of fast 
discs to store 
intermediary results being processed subsequently. 

For general values of the Mellin variable $N$ at three-loop order in Quantum Chromodynamics topologies contribute
also, for which root-valued letters appear in the alphabet. If these can be traded for the argument of the 
hyperlogarithm, the method remains applicable. This is, however, not the case for all massive 
3--loop topologies.
On the other hand, a remarkably wide class of diagrams can be calculated using the method of hyperlogarithms.
At the technical side the operator insertions are mapped to propagator-type factors referring to a representation
in terms of generating functions. At the end of the calculation the $N$th expansion coefficient has to be 
determined analytically for which techniques are available in the {\tt Mathematica}-package {\tt 
HarmonicSums}.
In some of the graphs multiply nested sums weighted by binomials of the type $\binom{2i}{i}$ in the numerator and 
denominator occur.
To construct the analytic continuation of these sums to $N \in \mathbb{C}$ their asymptotic expansion for
$|N| \rightarrow \infty, {\rm arg}(N) \neq \pi$ has to be calculated analytically. This requires the analytic 
Mellin-inversion
of the corresponding sum expressions. We used Risch-algorithm methods to compute the corresponding iterated
integrals, which request a larger amount of root-valued letters, cf. Ref.~\cite{BINOM} for details. Also a series 
of new special constants beyond those of the multiple zeta values and their cyclotomic and 
generalized sum generalizations emerges in these expressions. Operating in difference fields and using the 
Risch-algorithm we arrive at minimal representations algebraically keeping only functions with relative 
transcendence to each other. The present methods also allow the representation of the integrals calculated
in the present paper in $x$-space. Detailed transformation algorithms and results are given in \cite{BINOM}.  

The present analysis deals with convergent Feynman integrals only, while most of the Feynman graphs 
exhibit poles in the dimensional parameter $\ep = D - 4$. The calculation also of these diagrams requires a 
suitable regularization to be carried out first and still needs a thorough algebraic implementation. The major 
limiting factor for a general application of the algorithm to massive problems, including local operator insertions,
at present consists in the emergence of root-valued letters already at intermediate steps of the algorithm. 
A thorough mathematical treatment of these structures may be the subject of future investigations.

\vspace{5mm}\noindent
{\bf Acknowledgment.}~
We would like to thank A.~Behring, F.~Brown, D.J.~Broadhurst and A.~De~Freitas for discussions, M.~Steinhauser 
for providing the 
code {\tt MATAD 3.0}, and  A.~Behring for technical checks of the formulae. 
The graphs in the present paper were drawn using {\tt Axodraw} \cite{Vermaseren:1994je}.
This work was supported in part by DFG
Sonderforschungsbereich Transregio 9, Computergest\"utzte Theoretische Teilchenphysik, Studienstiftung des
Deutschen Volkes, the Austrian Science Fund (FWF) grants P20347-N18 and SFB F50 (F5009-N15), the European
Commission through contract PITN-GA-2010-264564 ({LHCPhenoNet}) and PITN-GA-2012-316704 ({HIGGSTOOLS}),
and by FP7 ERC Starting Grant  257638 PAGAP.

%-----------------------------------------------------------------------------

%-----------------------------------------------------------------------------
\end{document}